\documentclass[aps,onecolumn,preprint,prd,superscriptaddress,nofootinbib]{revtex4-1}
\usepackage{graphicx}
\usepackage{amsmath,amssymb,color,mathrsfs,verbatim,psfig,bbm,wasysym, pstricks,epsfig,slashed}

\newcommand{\beq}{\begin{equation}}
\newcommand{\eeq}{\end{equation}}
\newcommand{\beqa}{\begin{eqnarray}}
\newcommand{\eeqa}{\end{eqnarray}}
\newcommand{\ba}{\begin{array}}
\newcommand{\ea}{\end{array}}

\newcommand{\cL}{{\cal L}}

\begin{document}

\begin{titlepage}
 
\thispagestyle{empty}

\begin{flushright}UMD-PP-011-017\\
V2
\end{flushright}
\vspace{0.2cm}
\begin{center} 
\vskip 2.0cm
{\LARGE \bf   SUSY, the Third Generation and the LHC}
\vskip 0.5cm

\vskip 1.0cm
{\large Christopher Brust$^{2,1}$, Andrey Katz$^{1,3}$,\\ Scott Lawrence$^1$, and Raman Sundrum$^1$}
\vskip 0.4cm
{\it $^1$Department of Physics, University of Maryland, College Park, MD 20742}\\
{\it $^2$Department of Physics and Astronomy, Johns Hopkins University, Baltimore, MD 21218}\\
{\it $^3$ Center for the Fundamental Laws of Nature,  
Jefferson Physical Laboratory\\ Harvard University, Cambridge, MA 02138}\\
\vskip 1.7cm
\end{center}

\noindent 
We develop a bottom-up approach to studying SUSY with light stops and
sbottoms, but with other squarks and sleptons heavy and beyond reach
of the LHC. We discuss the range of
squark, gaugino and Higgsino masses for which the electroweak scale is radiatively
stable over the ``little hierarchy'' below  10 TeV. We review and
expand on indirect constraints on this scenario, in particular from
flavor and CP tests. We emphasize that in this context, R-parity violation is
very well motivated. The phenomenological differences between Majorana
and Dirac gauginos are also discussed. Finally, we focus on 
 the light subsystem of stops, sbottom  and neutralino with R-parity, in order to probe 
the current collider bounds.  We find that  1/fb LHC bounds are mild
and large parts of the motivated parameter space remain open, while
the 10/fb data can be much more decisive.

\end{titlepage}

\tableofcontents

\section{Introduction}
\label{sec:intro}

Supersymmetry (SUSY) remains a strong contender for the mechanism
underlying electroweak stability.  If one puts great stock in 
a particular high-energy SUSY model,  the model couplings can be run 
down and matched onto an 
LHC-energy effective Lagrangian, ${\cal L}_{eff}$, which can then be
used to carefully
 tailor experimental searches.
However, as is becoming increasingly clear, SUSY is a broad paradigm 
with several possible motivated incarnations and a complex parameter space,
and it is a challenge for experiments to
cover all of the
phenomenological bases. One way to proceed is to try to constrain
the form of ${\cal L}_{eff}$ by more bottom-up criteria, and to
use the results to guide experiments, committing perhaps to some broad
UV principles but not committing to a specific UV model. 

The most important such criterion is that  SUSY-breaking 
in ${\cal L}_{eff}$ be compatible with the radiative
stability of the electroweak scale  within the domain of validity of the
effective theory, up to roughly $10$ TeV.
The significance of the $10$ TeV scale is that almost all experiments, 
up to and including the LHC, only have
sensitivity to new physics $\lesssim 10$ TeV, be it through direct
searches or virtual effects. (In this regard, flavor physics tests are exceptional
in probing vastly higher scales and consequently they require special consideration.) 
The fact that the non-supersymmetric Standard Model (SM) is already fine-tuned in this
regime is known as the  ``little hierarchy problem'', and provides the
most immediate motivation for new physics accessible to colliders.

We must further weigh the
relevance for the effective theory of other 
general concerns of the SUSY paradigm, which at least partly 
relate to very high energies:
\begin{itemize}

\item  the SUSY Flavor Problem

\item  Grand Unification

\item  proton stability and R-parity

\item  superpartner dark matter candidates

\item  SUSY-breaking dynamics

\item Higgs mass

\end{itemize}

In this paper, we will focus on the 
 {\it minimal} effective theories that arise from the above
 viewpoint. They are ``minimal'' in terms of the particle content
 and parameter space of ${\cal L}_{eff}$. This does not  imply,
 however,
 that their UV-completions,  above LHC energies,
 are also minimal in some way. Conversely,  the Minimal
 Supersymmetric Standard Model (MSSM), is a minimal visible sector
 from the high-energy perspective, but is non-minimal in the sense
 that matters to the LHC effective theory and phenomenology, as we
 will review.
The central observation, mentioned in~\cite{Dimopoulos:1981zb,Sakai:1981gr} and developed in~\cite{Dimopoulos:1995mi}, 
is that radiative stability between the weak scale and $\sim 10$
TeV, does not require a superpartner in the effective theory for every
Standard Model (SM) particle, but just for those particles with 
order one couplings to the Higgs boson and electroweak breaking.
In this way, the minimal superpartner content is given by the gauginos,
Higgsinos, stop and sbottom, 
without sleptons or first and second generation squarks.\footnote{We 
will be more precise later.} 
The omitted superpartners
 may have masses above LHC reach
and may play a crucial role in weak scale stability up to much higher
inscales, but all this is outside the scope of the effective
theory and outside the grasp of the LHC.
Ref.~\cite{Cohen:1996vb} dubbed this kind of
structure, ``Effective Supersymmetry''. 
Since~\cite{Dimopoulos:1995mi,Cohen:1996vb}, a number of quite different approaches to far-UV
dynamics have converged on such a ``more minimal'' spectrum at
accessible energies~\cite{Sundrum:2009gv,Barbieri:2010pd,Craig:2011yk,Jeong:2011en}.

Of course, there is no
 guarantee that  at accessible energies
  new SUSY physics will be turn out to be minimal. 
Rather, we study minimal LHC-effective theories for three reasons:
(1) they represent possible SUSY
phenomenology, and there do exist UV SUSY
dynamics that match onto them, (2)  a great
deal of the natural parameter space remains open after one year of LHC
data, and yet discoverable within the next year, 
 (3) minimal models in any arena of exploration represent an important departure
point for thinking more broadly.

In this paper, we will take a more UV-agnostic
approach to the minimal effective theory at LHC energies than has been
previously considered.
We do not do this blindly, 
but only after discussion of the general SUSY concerns listed above. We
will argue that 
modern developments in
model-building and SUSY field theory have proliferated the range of
UV options that relate to these issues, and it is precisely for this
reason that we advocate thinking more modularly about them, and with
less commitment to any one UV plot.
Our goal will be to use
electroweak naturalness, flavor constraints, minimality, and earlier
searches
 as a guide to the LHC
phenomenology,  to discuss qualitative options (such as R-parity versus R-parity
violation) and to organize the different possible channels and
relevant parameter spaces. We will use this
platform to study the LHC phenomenology in more
detail, and in future work to broaden and help optimize experimental 
search strategies. We will adopt the name ``Effective SUSY'' to refer to
this minimalist and UV-agnostic approach to the LHC-effective
theory. Our study of effective SUSY coincides with the accumulation of significant LHC data. However, there are earlier collider studies relevant to effective SUSY on which our work expands, such as~\cite{Baer:1994xr,Baer:1998kt,Baer:2010ny,Kane:2011zd,Alwall:2010jc,Alwall:2011zm,Bhattacharyya:2011se}.

The paper is organized as follows. In Section \ref{sec:10tevlight}, we derive the
minimal effective SUSY Lagrangian subject to electroweak naturalness
with a cutoff of $\sim 10$ TeV. Here we impose R-parity and make
 the useful idealization that  the third generation does not mix with
 the first two generations. We also make the standard assumption that
 the Higgsino mass arises from a supersymmetric $\mu$ term. 
In Section \ref{sec:1tevlight}, we perform the same exercise but with a cutoff of only
$\sim 1$ TeV, in a sense increasing our agnosticism towards what lies
above the early 7 TeV LHC reach. One possibility, but not the only
one, is that this $1$ TeV effective theory derives straightforwardly
from the $10$ TeV effective theory of Section \ref{sec:10tevlight}. In Section \ref{sec:effsusyheavy}, we
study the possibility that Higgsinos obtain mass from soft SUSY
breaking rather than a $\mu$ term, and we write an even more
minimal set of effective Lagrangians with $10$ TeV and $1$ TeV
cutoffs. In Section \ref{sec:flavor} we put back consideration of
third-generation mixing, and
review and extend the constraints provided by
low-energy flavor and CP tests. We emphasize the considerable safety of the
effective SUSY scenario. In Section \ref{sec:rpv}, we make the case for R-parity
violation as a very plausible option, write the effective SUSY R-parity
violating interactions, and discuss some of the low-energy
constraints. In Section \ref{sec:dirac}, we discuss the interesting possibility of Dirac
gauginos and how this can considerably affect the collider phenomenology and
low-energy constraints. Section \ref{sec:pheno} is devoted to discussing collider
phenomenology, in particular the 7 TeV LHC. As a first foray, we focus
mostly on the minimal subsystem of stops, sbottoms and neutralino with
R-parity. We also make brief remarks about other phenomenological 
regimes of effective SUSY. Section IX provides our outlook.

While this paper was being completed, we became aware of three other
groups pursuing partially overlapping work~\cite{Kats:2011qh,Essig:2011qg,Papucci:2011wy}. 


\section{Effective SUSY $\lesssim 10$ TeV} 
\label{sec:10tevlight}
   
Let us start with the MSSM field content and ask which superpartners are
minimally needed in order to maintain electroweak naturalness below
 $10$ TeV,  roughly the collider reach in the years to come. We will
 not ask here what physics lies above this scale. Therefore at the technical
 level, $\Lambda_{UV} \equiv 10$ TeV provides the cutoff for
 any UV divergences encountered in the effective theory, and this allows us to
 estimate electroweak fine-tuning and check where in parameter space
 effective SUSY solves the ``little hierarchy problem'' of the SM. 

 SM particles with
 order one couplings to the Higgs boson must certainly have
 superpartners in the effective theory because they would otherwise
 give rise to quadratically divergent Higgs mass-squared
 contributions at one loop,  $\sim \Lambda_{UV}^2/(16 \pi^2)$, big enough to
 require significant fine-tuning. In order to supersymmetrically
 cancel these divergences, 
 the effective theory must therefore
 include the left-handed top and bottom squarks, $\tilde{q}_L \equiv (\tilde{t}_L,
 \tilde{b}_L)$, 
and the right-handed top squark, $\tilde{t}_R$,
 as well as the up-type Higgsino,
 $\tilde{h}_{u}$, and electroweak
 gauginos, $\lambda_{1, 2}$. 

Considerations beyond SUSY itself imply that we need to retain even 
more superpartners.
Electroweak gauge anomaly cancellation implies that  $\tilde{h}_u$
must be accompanied by $\tilde{h}_d$ in the effective theory. 
Indeed, one might have anticipated that  down-type Higgs {\it bosons},
$h_d$, are required anyway to give masses to
the down-type fermions, and that $\tilde{h}_d$ provide the required
superpartners.\footnote{We proceed with this logic in this section, although there
is a loop-hole  whereby 
$h_u$ can provide  down-type fermion masses in the effective theory,
and $h_d$ bosons are not needed. We discuss this option in Section \ref{sec:effsusyheavy}.}
With the $h_d$
bosons present in the effective theory, there is a new quadratic
divergence, even in the supersymmetric limit, in the form
of a (supersymmetric) hypercharge $D$-term. It is associated
by supersymmetry with the mixed hypercharge-gravity triangle anomaly. 
The quadratic divergence
vanishes only if Tr$(Y)=0$, where $Y$ is the hypercharge charge
matrix over the scalar fields of the effective theory. With the field
content described, including $h_d$, this condition is not satisfied,
and the theory remains unnatural despite superpartners for the main
players in the SM. Vanishing Tr$(Y)$ can be arranged by retaining the right-handed
bottom squark, $\tilde{b}_R^c$, within the effective theory. 

For the most part, two-loop quadratic divergences $\sim
\Lambda_{UV}^2/(16 \pi^2)^2$
 are not important
for Higgs
naturalness, with a cutoff as low as
$10$ TeV. But the QCD coupling is an exception. In particular, the
$\tilde{q}_L, \tilde{t}^c_R$ masses must themselves be so light in order to protect
Higgs naturalness at one loop order, that they suffer from their own
naturalness problem due to one-loop mass corrections from QCD. This
one loop QCD destabilization of the squarks, hence two-loop
destabilization of the Higgs, requires the gluino, $\lambda_3$, to be in the
effective theory.

In this way, the effective theory has complete
supermultiplets, 

\begin{align}
\label{eqn:superfields}
Q \equiv \left(\begin{array}{c}T \\ B\end{array}\right) &\equiv (\tilde{q}_L, q_L) \equiv \left(\left(\begin{array}{c}\tilde{t}_L \\ \tilde{b}_L\end{array}\right), \left(\begin{array}{c}t_L \\ b_L\end{array}\right)\right)\nonumber \\
\bar{T} &\equiv (\tilde{t}_R^c, t_R^c) \nonumber \\
\bar{B} &\equiv (\tilde{b}^c_R, b^c_R) \nonumber \\
H_u &\equiv (h_u, \tilde{h}_u)  \nonumber \\
H_d &\equiv (h_d, \tilde{h}_d) \nonumber \\
V_{1} &\equiv (B_{\mu}, \lambda_1) \nonumber \\
V_2 &\equiv (W_{\mu}, \lambda_2) \nonumber \\
V_3 &\equiv (G_{\mu}, \lambda_3) 
\end{align}
where we use the lower case ``$h$'' to distinguish
 just the scalars of the Higgs chiral supermultiplet,  ``$H$''.

\subsection{Effective Lagrangian, neglecting third-generation  mixing}
\label{sec:10tevlightlagrangian}

Above, we have introduced squarks belonging to only the ``third
generation'', 
and yet this notion is slightly ambiguous because generation-numbers are
not conserved, even in the SM. However, CKM mixing involving the third
generation is at least highly suppressed, so we will begin by considering the
``zeroth order'' approximation in which third-generation number is
exactly conserved. For most purposes in LHC studies of the new physics,
this approximation is sufficient. But for complete realism and to
check the viability of the theory
in the face of very sensitive low-energy flavor constraints, the extra subtlety of
third-generation mixing must be taken into account. We defer this
discussion until Section \ref{sec:flavor}. For now, this mixing is formally
``switched off''. Further, we will impose R-parity on effective SUSY,
and defer  the discussion of possible R-parity violating (RPV) couplings to Section \ref{sec:rpv}. 

 With the field content described above, the 
effective Lagrangian is given by
\begin{eqnarray}
\label{eqn:10tevlight}
{\cal L}_{eff} &=& \int d^4 \theta K + \left(\int  d^2 \theta
  \left(\frac{1}{4}{\cal W}_{\alpha}^2 + y_t \bar{T} H_u Q +  y_b
    \bar{B} H_d Q + \mu H_u H_d \right)+{\rm h.c.}\right) 
\nonumber \\
&~& 
+ {\cal L}_{kin}^{light}
-  \left(\bar{u}_R Y_{u}^{light} h_u \psi_L
+ \bar{d}_R Y_{d}^{light} h_d \psi_L+{\rm h.c.} \right)+ {\cal L}_{lepton} \nonumber \\
&~&  - m_{\tilde{q}_L}^2 |\tilde{q}_L|^2  - m_{\tilde{t}^c_R}^2 |\tilde{t}^c_R|^2 - m_{\tilde{b}^c_R}^2 |\tilde{b}^c_R|^2  -m_{h_u}^2 |h_u|^2  - m_{h_d}^2 |h_d|^2 \nonumber \\ &~&- \left(m_{i =1,2,3} \lambda_i \lambda_i + B\mu h_u h_d + A_t \tilde{t}^c_R h_u \tilde{q}_L + A_b \tilde{b}^c_R h_d \tilde{q}_L+{\rm h.c.}\right)\nonumber \\
&~& +{\cal L}_{hard} + {\cal L}_{non{\rm -}ren.},
\end{eqnarray}
where the first line is in superspace/superfield notation, while the
remaining lines are in components. Here, $K$ is the standard
gauge-invariant K\"{a}hler potential for the chiral superfields of
Eq. (\ref{eqn:superfields}), and ${\cal L}_{kin}^{light}$ denotes the standard gauge-invariant
kinetic terms for the light SM quarks  (that is, not the top and
bottom), $u_R, d_R, \psi_L \equiv (u_L,
d_L)$. ${\cal L}_{lepton}$ denotes all terms involving leptons, with Yukawa couplings to $h_d$ (neglecting neutrino mass terms). 
The super-field strength tensors are implicitly summed over all three gauge groups of the standard model, both here and throughout the paper.
Even the second line can be thought of as the result of  starting from the
supersymmetric MSSM, but then deleting all superpartners for light SM
fermions. As mentioned above, 
 we ignore the third generation mixing with the first two generations
(until Section \ref{sec:flavor}). 
The third and fourth lines are soft SUSY breaking terms for the
superfields of the effective theory. 

The absence of superpartners for
the light fermions will necessarily induce {\it hard} SUSY-breaking
divergences at one-loop order. To renormalize these,  we must
include hard SUSY breaking couplings into the effective Lagrangian,
and naturalness dictates that the renormalized couplings be at least
of one-loop strength, $\gtrsim 1/(16 \pi^2)$. These couplings are
included in the last line, in ${\cal L}_{hard}$. 
Such couplings can then 
appear within one-loop Higgs self-energy diagrams, yielding  two-loop sized
quadratic divergences, $\gtrsim \Lambda^2_{UV}/(16 \pi^2)^2$. While this
is acceptable from the viewpoint of naturalness, 
we see that we cannot tolerate order one hard breaking couplings. UV
completions of effective SUSY theory can contain mechanisms to naturally yield such
non-vanishing, but suppressed, hard breaking terms, for example~\cite{Sundrum:2009gv,Craig:2011yk}. Because the hard breaking is necessarily small, it is
largely negligible for early LHC phenomenology. On the other hand, at
a later stage of exploration, measuring hard SUSY breaking such as a difference
 between gauge and gaugino couplings may provide a valuable
 diagnostic.  

Effective SUSY is expected to arise from integrating out heavy physics
above 10 TeV, some of which is crucial in solving the hierarchy
problem to much higher scales. It should therefore be a
non-renormalizable effective theory, with higher-dimension
interactions suppressed by $\sim 10$ TeV or more. These are contained
in ${\cal L}_{non{\rm -}ren}$ on the last line. Again, these will be largely
irrelevant for early LHC phenomenology, but can very important in
precision low-energy experiments, such as CP or flavor tests. The
most stringent of such tests imply that at least some non-renormalizable
interactions have to suppressed by effective scales much beyond $10$ TeV. Again,
there are UV completions of effective SUSY which possess natural
mechanisms to explain this required structure. 

\subsection{Higgs mass}
\label{sec:10tevlighthiggsmass}

The experimental bounds on the lightest {\it physical} neutral Higgs scalar
provide some of the most stringent constraints on weak scale SUSY. 
The dominant couplings of our effective Lagrangian are just
those of the MSSM, so  the electroweak symmetry-breaking and 
Higgs-mass predictions are
essentially the same. This is problematic because naturalness dictates
stops lighter than a few hundred GeV, while the physical Higgs mass
constraints require higher stop masses. One difference with the
high-scale MSSM is that in effective SUSY we have
 hard SUSY breaking couplings, among which can be
 Higgs quartic couplings which ultimately contribute to the
physical Higgs mass. However, these contributions are modest, just a
few GeV, since the
hard SUSY-breaking couplings must be suppressed for electroweak naturalness. 
Instead, sizeable upward contributions to physical Higgs mass require
new particle content beyond the MSSM (see e.g.~\cite{Dine:2007xi} and references therein). For example, this is readily
accomplished by adding a chiral superfield gauge singlet to the effective
theory~\cite{Cavicchia:2007dp,Delgado:2010uj,Ross:2011xv}, 
\begin{eqnarray}
\delta {\cal L}_{eff} &=& \int d^4 \theta |S|^2 + \int d^2 \theta \left(\kappa
SH_u H_d +\frac{1}{2}\sigma S^2\right)  + {\rm h.c.} \nonumber \\  
&~&-m_s^2|s|^2 + {~\rm other~soft~terms}
\end{eqnarray} 
which contains a new contribution to the Higgs quartic
couplings, $\sim \kappa^2$. The soft scalar mass-squared term $m_s^2$
can be $O({\rm TeV}^2)$ without destabilizing EWSB. It can also ensure
that the singlet does not acquire a vacuum expectation.
In principle, in effective SUSY with a $10$ TeV cutoff, we must commit
to which type of physics, $\delta {\cal L}_{eff}$,
 accounts for an acceptable physical Higgs
mass. But for early LHC superpartner searches, the details of $\delta
{\cal L}_{eff}$ need not be relevant, as the new particles can lie
above $1$ TeV. In such cases, the new physics is just a 
 ``black box'' which gives viable physical Higgs masses. Indeed, 
in writing effective SUSY theories with a lower $\sim 1$ TeV cutoff,
we will see that we can formally imagine having integrated out the new
physics responsible for new Higgs quartic couplings.

\subsection{Naturalness in effective SUSY}
\label{sec:10tevlightnaturalness}

Here, we assemble the electroweak naturalness constraints on effective
SUSY, thereby giving a rough idea of the motivated regions of its parameter
space. For this purpose, we will compute various independent corrections to the $h_u$ mass-squared, and simply ask them to be $\lesssim (200 {\rm ~GeV})^2$ for
naturalness. We will compute these corrections {\it before} EWSB. Contributions sensitive to EWSB are typically $\sim O((100 {\rm ~GeV})^2)$, and therefore typically do not compromise naturalness. Given the intrinsically crude nature of naturalness
arguments, we see no merit in a more refined analysis.

We begin with a classical ``tuning'' issue. 
The $\mu$ term gives a supersymmetric $|\mu|^2$ contribution to
the Higgs mass-squareds. While the soft terms also contribute to Higgs
mass-squareds, naturalness forbids any fine cancellations, so
therefore by the criterion stated above,
\begin{equation}
|\mu| \lesssim
200 {\rm ~GeV}.
\end{equation} 
This same parameter then also plays the role of the Higgsino mass
parameter, ensuring relatively light charginos and
neutralinos in the superpartner  spectrum. (Of course, after EWSB, these physical states may
also contain admixtures of electroweak gauginos.) 

\begin{figure}[ht] 
 \centering
\includegraphics[width=6cm]{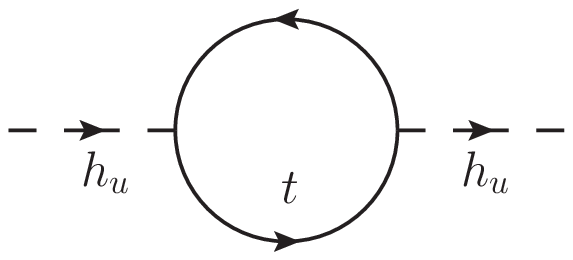}
\includegraphics[width=5cm]{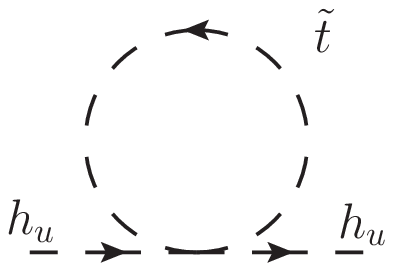}
\caption{Higgs mass corrections}
\label{fig:higgsmasst}
\end{figure}

Next, we turn to quantum loops. 
We assume that $\tilde{q}_L, \tilde{t}_R$ have approximately the 
same mass, $m_{\tilde{t}}$,  for simplicity, and we also neglect the
$\mu$ and $A$-terms. 
We work pre-EWSB since we are concerned with sensitivity to
parametrically higher scales. By evaluating the diagrams in figure \ref{fig:higgsmasst}, we find that the $m_{h_u}^2$ parameter receives the following correction:

\begin{equation}\label{eqn:higgsmass}
\delta m_{h_u}^2 = -\frac{3 y_t^2}{4\pi^2}m_{\tilde{t}}^2\ln \left(\frac{\Lambda_{UV}}{m_{\tilde{t}}}\right)\end{equation}

Naturalness therefore requires, very roughly, 
\begin{equation}
m_{\tilde{t}} \lesssim 400 {\rm GeV}.
\end{equation} 

\begin{figure}[ht] 
 \centering
\includegraphics[width=4.0cm]{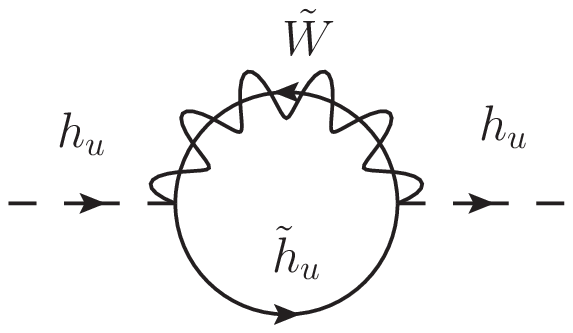}
\includegraphics[width=4.0cm]{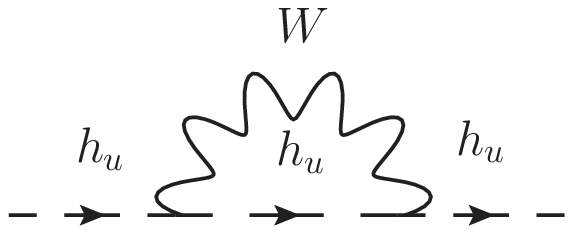}
\includegraphics[width=3.5cm]{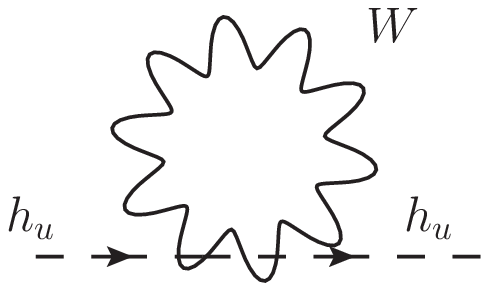}
\includegraphics[width=3.5cm]{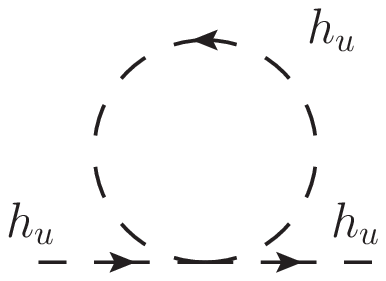}
\caption{Higgs mass correction}
\label{fig:higgsmassW}
\end{figure}

There are also electroweak gauge/gaugino/Higgsino one-loop
contributions to Higgs mass-squared. Again, working before electroweak
symmetry breaking (gaugino-Higgsino mixing) and just looking at the
stronger $SU(2)_L$ coupling, the Higgs
self-energy  diagrams are
in figure \ref{fig:higgsmassW}.

The Higgs mass correction is then given by
\begin{equation}
\label{eqn:higgsmassew}
\delta m_{h_u}^2=\frac{3g^2}{8\pi^2}(m_{\tilde{W}}^2+ m_{\tilde{h}}^2)\ln \frac{\Lambda_{UV}}{m_{\tilde{W}}}.\end{equation}

We identify
the Higgsino mass with $\mu$. Because we are already taking $\mu
\lesssim 200$ GeV, this translates into a roughly natural wino mass range of
\begin{equation}
m_{\tilde{W}} \lesssim {\rm TeV}. 
\end{equation}

Next, we compute the hypercharge $D$-term loop contribution to Higgs mass-squared, in figure \ref{fig:dterm}:

\begin{figure}[ht] 
 \centering
\includegraphics[width=6.0cm]{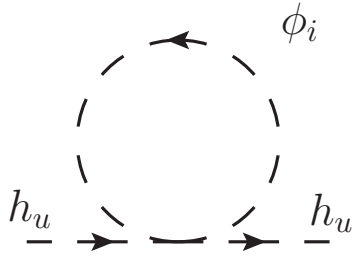}
\caption{Higgs mass correction}
\label{fig:dterm}
\end{figure}

This gives rise to a higgs mass correction:
\begin{equation}\delta m_{h_u}^2=\sum_{{\rm scalars}~i}\frac{{g'}^2Y_iY_{h_u}}{16\pi^2}\left(\Lambda_{UV}^2-m_i^2\ln\frac{\Lambda_{UV}^2+m_i^2}{m_i^2}\right).\end{equation}

Including both the right-handed sbottom and the down-type higgs, as we
do in this section, ensures that the quadratic divergence cancels, but
there is still a residual
correction to the higgs mass. Given that other scalars have already
been argued to be relatively light, we can use this correction to
estimate the natural range for the mass of $\tilde{b}_R$,
\begin{equation}
m_{\tilde{b}_R} \lesssim 3 {\rm TeV}.
\end{equation}

Finally, $\tilde{q}_L, \tilde{t}_R$ also being relatively light
scalars, suffer from their 
own naturalness problem, with mass corrections dominated by the
diagrams 
 in figure \ref{fig:stopmass}:

\begin{figure}[ht] 
 \centering
\includegraphics[width=4.0cm]{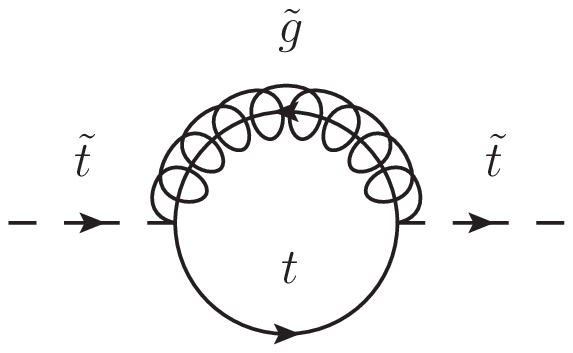}
\includegraphics[width=4.0cm]{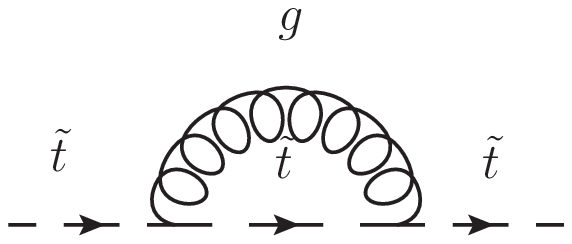}
\includegraphics[width=3.5cm]{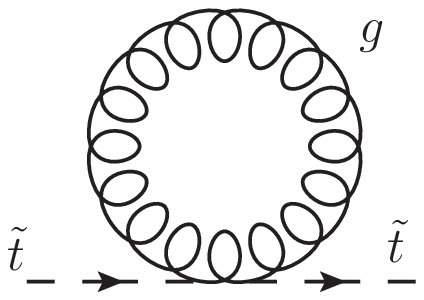}
\includegraphics[width=3.5cm]{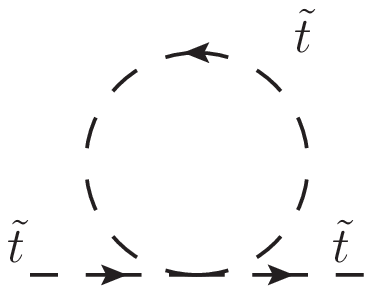}
\caption{Stop mass correction}
\label{fig:stopmass}
\end{figure}

This gives rise to a stop mass correction:
\begin{equation}\label{eqn:stopmass}\delta m_{\tilde{t}}^2 = \frac{2g_s^2}{3\pi^2}m_{\tilde{g}}^2\ln \frac{\Lambda_{UV}}{m_{\tilde{g}}}.\end{equation}

For squark masses $\sim$ few hundred GeV, naturalness requires
\begin{equation}
m_{\tilde{g}} \lesssim 2 m_{\tilde{t}}.
\end{equation}

\section{Effective SUSY $\lesssim 1$ TeV}
\label{sec:1tevlight}

Although the LHC has a multi-TeV reach in principle, parton
distribution functions fall so rapidly at high energies that most parton
collisions have sub-TeV momentum transfers. In the early LHC era, 
statistically signficant effective SUSY signals would be in this
regime. For example, in effective SUSY,
 gluino production would have a cross-section of just a few fb for TeV
 gluino mass. We can
therefore focus our attention on just the early accessible physics by constructing
 a rough effective SUSY theory with a cutoff
$\Lambda_{UV} \sim$ TeV, while not committing to the physics above this
scale. With such a low cutoff, only top quark loops in the SM
destabilize Higgs naturalness. This is cured by SUSY cancellation upon
including the squarks, $\tilde{q}_L, \tilde{t}_R$, to form complete
supermultiplets, $Q \equiv (\tilde{q}_L, q_L), \bar{T} \equiv
(\tilde{t}_R^c, t_R^c)$, as before. Even  hypercharge $D$-term
divergences from the uncanceled Tr$(Y)$ are not quantitively
significant. It therefore appears that we can dispense with Higgsinos,
$\tilde{b}_R$, and the gauginos in the effective theory. However, if
Higgsino mass arises from a supersymmetric $\mu$ term, as discussed in
Subsection \ref{sec:10tevlightnaturalness}, then electroweak naturalness also forces
the Higgsinos to be light. We will continue with this assumption in
this section, and therefore retain complete supermultiplets, $H_{u,d}
\equiv (h_{u,d}, \tilde{h}_{u,d})$.

Even though we do not commit here to the structure of the theory above
$1$ TeV, one possibility is that it is just that of the last
section. But in that case, by Eq. (\ref{eqn:stopmass}), we should include the gluino in
the sub-TeV effective theory. However, non-minimal physics in the $1
- 10$ TeV window can change this conclusion, and indeed the gluino
might naturally be considerably heavier than $1$ TeV. We illustrate
such new physics in Section \ref{sec:dirac}, with the example of a Dirac
gluino. It exemplifies the general theme that non-minimal UV physics
can lead to more minimal IR physics, while still being compatible with naturalness.
Here, we merely check within the TeV effective theory 
that naturalness  indeed requires
stops, but that these stops {\it do not
require} gluinos.  The first statement follows from Eq. (\ref{eqn:higgsmass}),  where 
naturalness up to $\Lambda_{UV} \sim 1$ TeV then implies
\begin{equation}
m_{\tilde{t}} \lesssim 700 {\rm GeV}.
\end{equation}
The second statement follows from Eq. (\ref{eqn:stopmass}), where we see that with the
logarithm of order one and gluino mass $\sim 1$ TeV, we can naturally
have stops as light as $300$ GeV. In our phenomenological studies of
Section \ref{sec:pheno}, we mostly keep in mind lighter stops, $m_{\tilde{t}}
\lesssim 400 {\rm GeV}$, compatible with either $1$ or $10$ TeV cutoffs as discussed
in Section \ref{sec:10tevlight}. 

\subsection{Effective Lagrangian, neglecting third-generation mixing}
\label{sec:1tevlight:lagrangian}

Given the light superpartner content described above, 
the R-parity conserving effective theory below a
TeV  is given by 
\begin{eqnarray}
\label{eqn:1tevlight}
{\cal L}_{eff} &=& \int d^4 \theta K + \left(\int  d^2 \theta
  \left(\frac{1}{4}{\cal W}_{\alpha}^2 + y_t \bar{T} H_u Q +  y_b
    \bar{B} H_d Q + \mu H_u H_d \right)+{\rm h.c.}\right) 
\nonumber \\
&~& 
+ {\cal L}_{kin}^{light}
-  \left(\bar{u}_R Y_{u}^{light} h_u \psi_L
+ \bar{d}_R Y_{d}^{light} h_d \psi_L+{\rm h.c.} \right)+ {\cal L}_{lepton} \nonumber \\
&~&  - m_{\tilde{q}_L}^2 |\tilde{q}_L|^2  - m_{\tilde{t}^c_R}^2 |\tilde{t}^c_R|^2  -m_{h_u}^2 |h_u|^2  - m_{h_d}^2 |h_d|^2 \nonumber \\ &~&-\left(B\mu h_u h_d +A_t \tilde{t}^c_R h_u \tilde{q}_L +{\rm h.c.}\right)\nonumber \\
&~& +{\cal L}_{hard} + {\cal L}_{non{\rm -}ren.},
\end{eqnarray}
which is to be interpreted as in
Eq. (\ref{eqn:10tevlight}) except that all {\it terms involving gauginos
or} $\tilde{b}_R^c$ {\it are to be thrown away} after expanding the
superspace expressions in components. 

With the cutoff as low as $1$ TeV, the hard SUSY-breaking can now
include $|h_u|^4$ couplings strong enough to give contributions to the
physical Higgs mass of tens of GeV without making EWSB scale
unnatural. One can think of these terms as arising from new fields,
 such as discussed in Section \ref{sec:10tevlighthiggsmass}, heavier than $1$ TeV, which have therefore been integrated out. One virtue of this sub-TeV theory is that we do not
have to commit to just what UV physics contributed to Higgs mass;
whatever it might be is parametrized by the effective hard couplings. 

\subsection{Dark Matter considerations}
\label{sec:1tevlight:dm}

In our TeV effective theory, we must take the Higgsinos as the
lightest superpartners in
order to avoid  phenomenologically dangerous colored (collider-)stable
particles in the form of stops or sbottom. Such Higgsinos will then
form charginos and neutralinos at the ends of superpartner decay
chains. Higgsino neutralinos would have a
thermal relic abundance smaller than needed to fully account for
all of dark matter. This is not an issue if dark matter is dominated
by other physics not (soon to be) accessible to the LHC. Another
possibility is that the wino and bino, $\lambda_{1,2}$, which are not
required to be light by naturalness, are nevertheless light and in the
effective theory, and a linear combination of gaugino-Higgsino forms a
neutralino LSP. It is possible then that such a hybrid LSP has the
correct thermal relic abundance to account for dark matter.  This
computation still remains to be checked in the effective SUSY context however.
 Even in this case, our minimal effective theory is still
useful, in that for the purposes of {\it early} LHC phenomenology the details of
charginos/neutralinos are not as important as their existence and the
LSP mass. The Higgsino LSP in our effective theory can therefore serve as a toy
model of whatever the real chargino/neutralino degrees of freedom
are. More refined modeling can wait until the new physics is
discovered. 

\section{Effective SUSY with heavy Higgsinos}
\label{sec:effsusyheavy}
\subsection{Effective SUSY with $10$ TeV cutoff}
\label{sec:10tevheavy}

As alluded to earlier, given that we necessarily have hard SUSY
breaking couplings in effective SUSY, 
we can  reduce the particle content even further by
eliminating $h_d$ {\it bosons} and the right-handed bottom squark $\tilde{b}^c_R$ from
the effective theory. See Refs~\cite{Dobrescu:2010mk,Ibe:2010ig,Davies:2011mp} for earlier related works. 
This move maintains the vanishing of Tr$(Y)$
required for naturalness with $10$ TeV cutoff, but forces us to obtain
Yukawa-couplings for
down-type fermions by coupling them to
\begin{equation}
h_u^* \equiv i\sigma_2 h_u^{\dagger},
\end{equation}
 where $\sigma_2$ is the second weak-isospin Pauli matrix. This is the usual approach
 to getting down-type fermion masses in the SM with a single Higgs
 doublet. In the SUSY context, such a coupling cannot arise from a
 superpotential, which can only depend on $H_u$, not
 $H_u^{\dagger}$. Instead, it represents a hard SUSY breaking effect 
(though it may arise from soft SUSY breaking
from the vantage of a UV completion).   It poses no threat to
naturalness if the couplings are $\ll 1$. This is certainly the
case for all the down-type Yukawa couplings.\newline 

\subsubsection{Effective  Lagrangian, neglecting third-generation mixing}
\label{sec:10tevheavylagrangian}

With the particle content described above, the R-parity conserving
effective Lagrangian is given by
\begin{eqnarray}
\label{eqn:10tevheavy}
{\cal L}_{eff} &=& \int d^4 \theta K+
\left(\int  d^2 \theta \left(\frac{1}{4}{\cal W}_{\alpha}^2 + y_t \bar{T} H_u Q \right)+{\rm h.c.}\right)\nonumber \\
&~& + {\cal L}_{kin} - \left(\bar{u} Y_{u}^{light} h_u q_L+y_b \bar{b} h_u^* q_L 
+ \bar{d} Y_{d}^{light} h_u^* q_L +{\rm h.c.}\right)+ {\cal L}_{lepton}
\nonumber \\
&~&  - m_{\tilde{q}_L}^2 |\tilde{q}_L|^2  - m_{\tilde{t}^c_R}^2 |\tilde{t}^c_R|^2  -m_{h_u}^2 |h_u|^2
\nonumber \\
&~& - \left(m_{i =1,2,3} \lambda_i \lambda_i + A \tilde{t}^c_R h_u \tilde{q}_L + m_{\tilde{h}} \tilde{h}_u \tilde{h}_d + {\rm h.c.}\right)
\nonumber \\
&~& + {\cal L}_{hard} + {\cal L}_{non{\rm -}ren.}.
\end{eqnarray}

The Kahler potential $K$ consists of the gauge-invariant kinetic terms
for the chiral superfields, $\bar{T}, Q, H_u$, while compared with Eq. (\ref{eqn:10tevlight}), the kinetic terms for the (now un-superpartnered) fermions $b_R$ and $\tilde{h}_d$ have now been added to 
 ${\cal L}_{kin}$. The second to fourth lines still follow from the
 MSSM after deleting fields that are absent in our effective theory,
 except  for the small Yukawa couplings of $h_u^*$ to down-type fermions, which
 we pointed out above are a form of hard SUSY breaking. Other hard
 breaking as well as non-renormalizable couplings appear on the last
 line. Our discussion of the physical Higgs mass, and contributions to
 it, is similar to subsubsection~\ref{sec:10tevlighthiggsmass}. However a singlet coupling to $h_u h_d$ is not possible since we have removed $h_d$, but in an electroweak triplet coupled to $H_u H_u$ is possible and results in a $|h_u|^4$ terms in the potential~\cite{Espinosa:1991gr,Espinosa:1992hp,Espinosa:1998re}. 

\subsubsection{Higgsino mass}
\label{sec:10tevheavyhiggsinomass}

Note that the Higgsino mass now takes the form of a soft SUSY-breaking
mass term, $m_{\tilde{h}}$,  as opposed to a supersymmetric $\mu$ term as in Section
\ref{sec:10tevlight}. In this way, it is uncorrelated with any contribution to Higgs
boson mass-squared. Therefore, there is only one modification to the bounds obtained in Section \ref{sec:10tevlightnaturalness}; namely, that now  $m_{\tilde{h}}$ is now only constrained by Eq. (\ref{eqn:higgsmassew}), so that
\begin{equation}
m_{\tilde{h}} \lesssim {\rm   TeV}.
\end{equation}

\subsection{Effective SUSY $\lesssim 1$ TeV}
\label{sec:1tevheavy}

In the most minimal of our effective theories, all gauginos and
Higgsinos can naturally be heavier than a TeV and thus integrated out
of the sub-TeV effective theory. If
we identify $h_u$ with the SM Higgs doublet, the only new particles
are $\tilde{t}_L, \tilde{b}_L, \tilde{t}_R$.  \newline

\subsubsection{Effective Lagrangian, neglecting third-generation mixing}
\label{sec:1tevheavylagrangian}

 The effective Lagrangian with R-parity is then given by
\begin{eqnarray}\label{eqn:1tevheavy}
{\cal L}_{eff} &=&\mathcal{L}_{SM}+\mathcal{L}_{kin}^{squarks} - V_{D-terms} - y_t^2(|h_u\tilde{q}_L|^2 + |\tilde{t}_R^c \tilde{q}_L|^2+ |\tilde{t}_R^c h_u|^2)\nonumber \\
&~& -m_{h_u}^2 |h_u|^2- m_{\tilde{q}_L}^2 |\tilde{q}_L|^2  - m_{\tilde{t}^c_R}^2 |\tilde{t}^c_R|^2  -\left( A \tilde{t}^c_R h_u \tilde{q}_L+{\rm h.c.}\right)
\nonumber \\
&~& + {\cal L}_{hard} + {\cal L}_{non{\rm -}ren.},
\end{eqnarray}

$\mathcal{L}_{SM}$ is the SM Lagrangian with $h_u$ playing the role of the SM Higgs doublet, but with no Higgs potential. The Higgs potential is a combination of the soft Higgs mass term in the second line, the $D$-term potential and possible hard SUSY-breaking couplings $\sim |h_u|^4$. As discussed in Subsection IIIA, these hard SUSY breaking couplings can be large enough to easily satisfy the Higgs mass bound without spoiling naturalness.



With exact R-parity, one of
the colored superpartners would necessarily be stable and
phenomenologically dangerous. However, we can use the above effective
Lagrangian as the minimal departure point for adding R-parity violating
corrections. We take this up in Section \ref{sec:rpv}. 

\subsubsection{Effective Lagrangian $\lesssim$ TeV, with
  neutralino LSP }
\label{sec:1tevheavylagrangianneutralinolsp}

Another possibility is that R-parity is exact but there is a 
neutralino LSP  in the spectrum, even
though it is not required by electroweak naturalness. It may or may
not be the dominant constituent of dark matter. Since we cannot
determine its identity by theoretical considerations alone, we will
just add a temporary ``place-holder'', that allows the squarks to
decay promptly while preserving R-parity. We choose this to be the
bino, $\lambda_1$, even though taken literally, it would predict too
large a thermal relic abundance of dark matter. A more refined
description of the neutralino would not add much to the early LHC
search strategy. In this option, as compared to that of Subsection \ref{sec:1tevlight:dm}
and Eq. (\ref{eqn:1tevlight}), we do not have a chargino. 

The effective Lagrangian then takes the form

\begin{eqnarray}\label{eqn:1tevlightneutralino}{\cal L}_{eff} &=&\mathcal{L}_{SM}+\mathcal{L}_{kin}^{squarks} - V_{D-terms} - y_t^2(|h_u\tilde{q}_L|^2 + |\tilde{t}_R^c \tilde{q}_L|^2+ |\tilde{t}_R^c h_u|^2)\nonumber \\
&~& -m_{h_u}^2 |h_u|^2- m_{\tilde{q}_L}^2 |\tilde{q}_L|^2  - m_{\tilde{t}^c_R}^2 |\tilde{t}^c_R|^2  -\left( A \tilde{t}^c_R h_u \tilde{q}_L+{\rm h.c.}\right)
\nonumber \\
&~& +i\bar{\lambda}_1\partial . \sigma\lambda_1 - \left(m_1\lambda_1\lambda_1 + {\rm h.c.}\right)-\sqrt{2}g'\left(\frac{1}{6}\bar{q}_L\bar{\lambda}_1\tilde{q}_L + \frac{1}{6}\bar{\tilde{q}}_L\lambda_1 q_L -\frac{2}{3} \bar{t}_R^c \bar{\lambda}_1\tilde{t}_R^c - \frac{2}{3}\bar{\tilde{t}}_R^c\lambda_1t_R^c\right) \nonumber \\
&~& + {\cal L}_{hard} + {\cal L}_{non{\rm -}ren.},
\end{eqnarray}

\section{Flavor-Changing Neutral Currents and CP Violation}
\label{sec:flavor}
Above, we have worked in the drastic approximation that the mixing between
the third generation with the first two generations  vanishes, so that
the meaning of ``third generation'' squarks, $\tilde{q}_L, \tilde{t}_R^c, \tilde{b}_R^c$, is
completely unambigous. In this limit, there is a conserved third-generation
(s)quark number. 
In the real world, third generation
mixing is non-zero but small. In Wolfenstein parametrization, mixing
with the second generation is of order $\epsilon^2$ and mixing with
the first generation is of order $\epsilon^3$, where $\epsilon \sim
0.22$ corresponds to Cabibbo mixing. 
Given this fact, it is more natural to have
 comparable levels of violation of third-generation
(s)quark number in the physics we have added beyond the SM.

In practice this means that for every interaction term in which the
squarks currently appear, where third-generation number is conserved
by the presence of  $t$ or $b$ quarks (in electroweak gauge basis), we now allow more
general couplings, with the third generation quarks replaced by quarks
of the first and second generations. The associated couplings with second
generation quarks are taken to be of order $\epsilon^2$, while those
with first generation quarks are taken to be of order $\epsilon^3$,
all in electroweak gauge basis. All these couplings involving the squarks are
technically hard breaking of SUSY, but $\epsilon^{2(3)}$ is
so small that, like other hard breaking in the effective
theory, they do not spoil Higgs naturalness below 10 TeV. 
For most, but not all, of the LHC collider phenomenology the small
$\epsilon^{2(3)}$ effects are negligible and we can proceed with our
earlier 
effective Lagrangians. (We must of course keep SM third generation mixing effects,
so that, for example, the bottom quark decays.)
But in the more realistic setting with third-generational mixing, we must
confront the SUSY flavor problem. In effective SUSY, this problem has
two faces, IR and UV.

The UV face of the problem is contained in the
non-renormalizable interactions of Eq.~\eqref{eqn:10tevlight}. For example, they can include
flavor-violating interactions such as $\bar{s} d \bar{s} d$. If such
a non-renormalizable interaction were suppressed only by $(10$
TeV$)^2$, it would lead to FCNCs in kaon mixing, orders of magnitude
greater than observed.  It is therefore vital for the
non-renormalizable interactions to have a much more benign flavor
structure. Whether this is the case or not is 
determined by matching to the full theory above 10 TeV, IR effective SUSY
considerations alone cannot decide the issue. Refs.~\cite{Sundrum:2009gv,Craig:2011yk} are examples of
UV theories which reduce to effective SUSY at accessible energies and
automatically come with the kind of benign UV flavor structure we
require. In this paper, we simply assume that the UV-sensitive non-renormalizable
interactions are sufficiently flavor-conserving to avoid conflict with
FCNC constraints. 

There remain FCNC effects that are UV-insensitive but are assembled in
the IR of the effective theory through the small $\epsilon^{2(3)}$
flavor-violating couplings. Many of these  have been studied in
Refs.~\cite{Giudice:2008uk} and are small enough to satisfy current
constraints. Indeed this feature is one of the selling points of
effective SUSY. Here, we illustrate one such FCNC effective
interaction for (CP-violating) $K-\bar{K}$ mixing arising as a SUSY
``box'' diagram, that has not yet been treated in the
literature. It is particularly dangerous because of the strong QCD
renormalization in running the effective operator induced by
a superpartner loop down to the hadronic scale.
While the effect is suppressed  by ${\cal
  O}(\epsilon^{10})$  in effective SUSY, it is more stringently
constraining than $B_d-\bar{B}_d$ mixing or $B_s-\bar{B}_s$ mixing,
even though these are suppressed by just ${\cal
  O}(\epsilon^6)$ and ${\cal
  O}(\epsilon^4)$ respectively. We
show that with our rough flavor-changing power-counting  the
$\tilde{b}_R$ squark is constrained to lie above several TeV. 

\begin{figure}[ht] 
 \centering
\includegraphics[width=7.0cm]{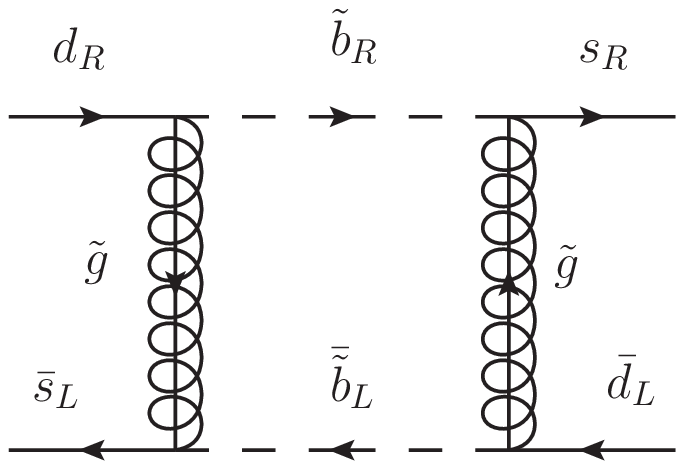}
\includegraphics[width=7.0cm]{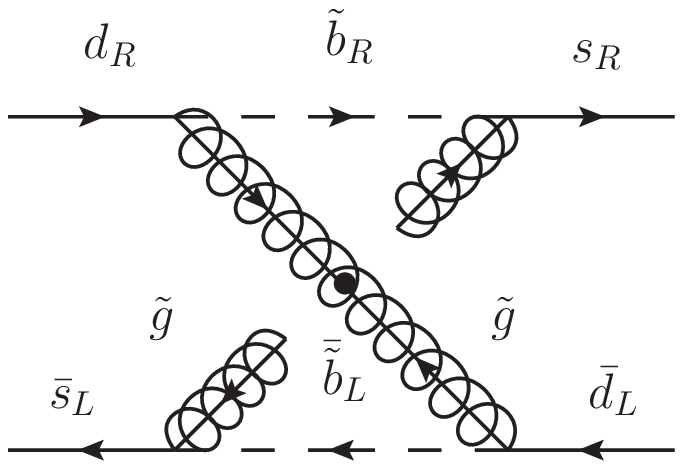}
\caption{Contributions to $K-\bar{K}$ mixing}
\label{fig:kkbarR}
\end{figure}

In a low-energy effective Lagrangian to be run down to the hadronic scale, we match onto the effective operator:

\begin{equation}\mathcal{L}_{eff}\supset \kappa (\bar{s}_Ld_R)(\bar{s}_Rd_L).\end{equation}
Integrating out the superpartners yields:

\begin{align}\kappa&\sim -\frac{g_s^4\epsilon^{10}}{4\pi^2}\frac{2}{3}\frac{m_3^2}{(m_3^2-m_{\tilde{q}_L}^2)^2(m_3^2-m_{\tilde{b}_R^c}^2)^2(m_{\tilde{q}_L}^2-m_{\tilde{b}_R^c}^2)}\nonumber \\ &\qquad \times ((m_{\tilde{q}_L}^2-m_{\tilde{b}_R^c}^2)(m_3^2-m_{\tilde{b}_R^c}^2)(m_3^2-m_{\tilde{q}_L}^2)+m_{\tilde{q}_L}^2(m_3^4+m_{\tilde{b}_R^c}^4)\ln\frac{m_{\tilde{q}_L}^2}{m_3^2}\nonumber \\ & \qquad+2m_3^2m_{\tilde{q}_L}^2m_{\tilde{b}_R^c}^2\ln\frac{m_{\tilde{b}_R^c}^2}{m_{\tilde{q}_L}^2}+m_{\tilde{b}_R^c}^2(m_{\tilde{q}_L}^4+m_3^4)\ln\frac{m_3^2}{m_{\tilde{b}_R^c}^2})\nonumber \\ &- \frac{g_s^4\epsilon^{10}}{8\pi^2} \frac{1}{12} \frac{1}{(m_3^2-m_{\tilde{q}_L}^2)^2(m_3^2-m_{\tilde{b}_R^c}^2)^2(m_{\tilde{q}_L}^2-m_{\tilde{b}_R^c}^2)}\nonumber \\ &\qquad \times ((2m_3^2m_{\tilde{q}_L}^4m_{\tilde{b}_R^c}^2 - m_3^4m_{\tilde{q}_L}^4)\ln\frac{m_3^2}{m_{\tilde{q}_L}^2}+(2m_3^2m_{\tilde{q}_L}^2m_{\tilde{b}_R^c}^4-m_3^4m_{\tilde{b}_R^c}^4)\ln\frac{m_{\tilde{b}_R^c}^2}{m_3^2}\nonumber \\ & \qquad +m_{\tilde{q}_L}^4m_{\tilde{b}_R^c}^4\ln\frac{m_{\tilde{q}_L}^2}{m_{\tilde{b}_R^c}^2} + m_3^2(m_3^2-m_{\tilde{q}_L}^2)(m_3^2-m_{\tilde{b}_R^c}^2)(m_{\tilde{q}_L}^2-m_{\tilde{b}_R^c}^2)),\end{align}
where, as discussed above, the squark couplings to second generation
quarks are assigned strength $\sim g_s \epsilon^2$, while squark
couplings to the first generation are $\sim g_s \epsilon^3$. We neglect $\tilde{b}_L$-$\tilde{b}_R$ mixing (after EWSB).

Current constraints on $\epsilon_K$ require that~\cite{Isidori:2010kg} 

\begin{equation}(\mathrm{Im}(\kappa)) \lesssim \left(\frac{1}{3 \times 10^5 \mathrm{~TeV}}\right)^2.\end{equation}
For $m_3 \sim$ TeV and $m_{\tilde{q}_L} \sim 350$ GeV, this translates into a
bound on $\tilde{b}_R$ mass of roughly $m_{\tilde{b}_R^c} \gtrsim 17 {\rm ~TeV}$.

Of course, this bound is extremely sensitive to our estimates for the
flavor-changing vertices. For example, if each flavor-changing vertex
were only half as strong as our above estimates, the bound would be
relaxed to $m_{\tilde{b}_R^c} \gtrsim 4 {\rm ~TeV}$, roughly
consistent with the requirements of naturalness in Subsection
\ref{sec:10tevlightnaturalness}. Alternatively, there may be small
phases present in the vertices that further suppress $\kappa$. In the even
more minimal 
 effective supersymmetry structure of Section \ref{sec:effsusyheavy},
  $\tilde{b}_R^c$ is completely absent and there is no robust infrared
  contribution to $\kappa$ at one-loop order to worry about.

There are also CP-violating effects unrelated to flavor-changing,
in particular electric dipole moment (EDM) constraints.  From~\cite{Hamzaoui:1997kt,Li:2010ax} (see also references therein),
we see that again effective SUSY has a relatively safe IR structure. 
The tightest constraints  come from the neutron EDM.\footnote{Here, we
  are discussing the supersymmetric CP problem as opposed to the
  Strong CP problem. We assume for concreteness that the Peccei-Quinn
  mechanism with an axion resolves the Strong CP problem.} 
 Two-loop effects
of stop, tops and gluinos match onto the Weinberg operator, which in
turn induces the neutron EDM. Even without cancelation among phases in soft
masses, sizeable but not arbitrary phases,
$\lesssim 1/3$ radian, are allowed for gluino mass $\sim$ TeV. We show
that these constraints are even more relaxed in the case of Dirac gauginos, in
Section \ref{sec:dirac}.

\section{R-Parity versus R-Parity Violation}
\label{sec:rpv}

R-parity plays a central role in theory and phenomenology within
 the weak scale SUSY paradigm. We will review some of the
 reasons for this, and argue
 that in light of several modern theoretical developments, the case
 for R-parity conservation in effective SUSY is less compelling.
 We are therefore more strongly motivated to take seriously an
 R-parity violating phenomenology. Quite apart from these theoretical
 considerations,
we believe that this RPV phenomenology of effective SUSY is quite
distinctive, and has so far received little attention. We will take up
its study in future work. 

\subsection{Proton decay}
\label{sec:rpvprotondecay}

The standard motivation for R-parity is that it leads to conserved
baryon  number. But it does not
follow in complete generality. In  the MSSM, 
 baryon-number conservation only
follows from R-parity after restricting to renormalizable
interactions. For example, R-parity conserving but 
non-renormalizable superpotential interactions of the general  form
$W \propto \bar{U} \bar{U} \bar{D} \bar{E}$ give rise to proton decay. 
If the MSSM is taken as
valid up to an extremely high scale, such a non-renormalizable term,
and the resulting proton decay rate, would be suppressed by that high
scale. However, if the MSSM is  an
effective theory emerging only below some lower threshold, 
then the non-renormalizable operator can be suppressed by just 
this lower threshold scale, leading to excessive proton
decay. This is precisely the issue in many SUSY GUT
theories, where such an effective interaction arises in the effective
MSSM after integrating out a color-triplet GUT-partner of the
Higgs. The moral only gains strength in
 effective SUSY, with  a 10 TeV cutoff.
For example, a dimension-6 R-parity
conserving operator
such as $u_Ld_Lu_Re_R$ can be viewed as a remnant of a supersymmetric
non-renormalizable Kahler potential term. It gives rise to extremely rapid
proton decay if suppressed by just $(10 {\rm~TeV})^2$.  
Such an operator might well arise upon integrating out
 new thresholds above $10$ TeV. 

We conclude that R-parity is not by itself 
enough to protect against proton decay in effective SUSY, in general we
need some other symmetry, such as baryon-number or lepton-number
symmetry.\footnote{While baryon number (lepton number)
  is broken by anomalies, just as in the SM, this need only imply baryon number
  violation  via non-perturbatively small
  interactions, which can easily be well below any experimental
  bounds.}    Clearly then, the proton-stability motivation for R-parity
is gone. 

\subsection{Unification}
\label{sec:rpvunification}

Traditionally, the reason for arguing against new physics thresholds
between the GUT and weak scales is because such new physics 
generally spoil the success of gauge coupling  unification. But this
is evaded if the new physics comes in complete GUT multiplets. 
For example, this is what is typically assumed for the messenger
threshold of gauge-mediated SUSY-breaking models.
 In 
the model-building of recent years, we have seen that even quite
radically new intermediate structure can maintain the success of gauge
coupling 
unification by following this basic rule of GUT-degenerate
thresholds~\cite{Pomarol:2000hp}. There also exist new unification mechanisms that improve
on the imperfect unification of SM via 
strong coupling effects over intermediate scales~\cite{Agashe:2005vg}.
Therefore, we cannot have confidence that there is a Weak-GUT desert,
as is often assumed. There may well be important new physics (not far)
above 10 TeV, and in this context R-parity does 
not save us from excessive proton decay, as discussed above.

Another GUT-related reason in favor of R-parity is that
in the context of traditional GUT models,
imposing  baryon- or lepton-number symmetry conflicts
with the unification of
quarks and leptons, whereas imposing R-parity does not. However, such
traditional GUT models also suffer from other difficulties such as the notorious
doublet-triplet splitting problem. In more recent years, it has been
understood that some of the successes attributed to SUSY GUTs can
arise more generally, in particular in the context of Orbifold GUT
models (see~\cite{Kawamura:2000ev,Hall:2002ea} and references therein). Such models employ  ``split multiplets'', in which quarks and
leptons can naturally arise as incomplete parts of {\it separate} GUT
multiplets, and the Higgs doublet and triplet are also neatly split in the same manner.
In this orbifold unification context, one can straightforwardly 
impose
baryon- or lepton-number symmetry, safeguarding proton stability
without requiring R-parity.

In this way, the unification considerations that originally favored
R-parity over baryon- or lepton-number  symmetry are less compelling. 

\subsection{Dark Matter}
\label{sec:rpvdm}

There is a second traditional motivation for R-parity, namely that
the lightest R-odd superpartner is stable, and therefore may account for
the dark matter of the Universe, enjoying the rough  quantitative
success known as the ``WIMP-miracle''. RPV interactions spoil this stability
and seem to rob us of such a dark matter candidate. However, it is
entirely possible that dark matter does consist of weak scale WIMPs,
but these WIMPs are stabilized by carrying a different symmetry than
R-parity, under which the SM is inert. This natural possibility leads us to separate the
question of modeling dark matter from the questions of electroweak and
Higgs naturalness, at least for the immediate purpose of pursuing
collider phenomenology. In the traditional view, every superpartner
produced cascade decays down to the dark matter particle. But more
generally, we can have R-parity violation and dark matter may or may
not be at the end of superpartner decay chains.

\subsection{RPV and FCNCs}
\label{sec:rpvfcnc}

A final reason for favoring R-parity is that in standard weak scale
SUSY, large parts of RPV parameter space lead to excessive FCNCs, only exacerbating
 the usual SUSY Flavor Problem. However, this point is mitigated,
 though not completely evaded, in effective SUSY, because of the
 greatly reduced squark content, as discussed below. Again, this makes
 RPV a more motivated possibility in the effective SUSY context. 

~

In the end, we think that both R-parity and RPV alternatives are
plausible in the effective SUSY context, and make for very
different phenomenological features and search strategies. Below we
discuss RPV with proton decay protected by lepton number symmetry, and
alternatively by baryon number conservation.

\subsection{RPV with Lepton number conservation}
\label{sec:rpvleptonnumber}

The standard renormalizable RPV SUSY couplings preserving lepton
number are of the superpotential form $W \propto \bar{U}_I \bar{D}_J \bar{D}_K$,
with generational indices $I,J,K$. 
Such couplings give rise to a variety of RPV Yukawa couplings and
(after SUSY breaking) RPV A-terms which can decisively affect  
superpartner decays and flavor physics. Here, we specialize to the most
minimal particle content of effective SUSY, as discussed in Section
\ref{sec:effsusyheavy},  with beyond-SM field content given by $\tilde{q}_L, \tilde{t}^c_R, \tilde{h}_{u,d},
\lambda_i$. While there is  the up-type scalar singlet $\tilde{t}^c_R$, there is
no down-type scalar singlet, and therefore no RPV A-terms are possible
in the effective SUSY theory. 
The only RPV Yukawa couplings that come from truncating the above type
of superpotential to effective SUSY are of the form
\begin{equation}
\label{eqn:rpvudd}
{\cal L}_{RPV} = \kappa_{IJ} \tilde{t}_R^c d_R^{cI} d_R^{cJ}.
\end{equation}
We will consider this to be added to the minimal 10 TeV effective Lagrangian
of Eq. (\ref{eqn:10tevheavy}), or the 1 TeV effective Lagrangian of Eq. (\ref{eqn:1tevheavy}). 

Flavor constraints on these couplings, reviewed in  Ref.~\cite{Barbier:2004ez}, 
easily allow RPV coupling strengths that lead to prompt squark
decays into quarks at colliders. But while lepton-number conservation is
sufficient to 
protect against
proton decay (assuming the gravitino or other non-minimal fermions are
heavier than the proton), it does not forbid neutron-antineutron
oscillations. This is because (accidental) $U(1)$ baryon-number symmetry is
incompatible with the combination of RPV couplings, gaugino-squark-quark coupling,
and Majorana gaugino masses. The bounds on neutron-antineutron
oscillations are stringent (see~\cite{Mohapatra:2009wp} for review), even in effective SUSY where CKM
suppressions are incurred in
mediating such effects via the third generation squarks and
gauginos. Again, RPV couplings can straightforwardly be strong enough
to lead to prompt squark decays to quarks at colliders. And yet, they
cannot be order one in strength. Theoretically, having RPV
couplings $\ll 1$ is plausible enough, related perhaps to the
smallness of ordinary Yukawa couplings. Experimentally, small RPV
couplings imply that squarks
cannot be {\it singly} produced at colliders. 

 Remarkably, there is a
way of recovering  $U(1)$ baryon number symmetry consistent 
with {\it order one} RPV couplings  of the form
of Eq. (\ref{eqn:rpvudd}), but it requires realizing gauginos as components of Dirac
fermions.  Observing
single squark production can then be an interesting diagnostic of
supersymmetry breaking, even those parts out of direct reach of
the 7 TeV LHC. We will show how this works in Section \ref{sec:dirac}.

\subsection{R-parity violation with Baryon number conservation}
\label{sec:rpvbaryonnumber}
The standard renormalizable RPV SUSY couplings preserving baryon
number are superpotential terms of the form, $W \sim LL \bar{E}, Q L
\bar{D}, LH_u$.  Let us again consider truncating to the minimal beyond-SM field
content described in Section \ref{sec:effsusyheavy},  $\tilde{q}_L, \tilde{t}^c_R, \tilde{h}_{u,d},
\lambda_i$.  Again, there are no A-terms of the forms of these
superpotentials possible, and
the $LL \bar{E}$ completely vanishes.
The
bilinear superpotential turns into a mixing mass term $\ell
\tilde{h}_u$. Since $\tilde{h}_d$ and the left-handed leptons, $\ell$,
share the same gauge quantum numbers, we can choose a new basis for
them such that there are no $\ell \tilde{h}_u$ terms.  The 
only surviving RPV Yukawa couplings are then of the form,
\begin{equation}
{\cal L}_{RPV} = \kappa'_{IJ} d_R^{cI} \ell_{L}^J \tilde{q}_L.
\end{equation}

We defer the study of the flavor constraints and the LHC implications of this
type of baryon-number conserving RPV interactions within effective
SUSY to future work.
Ref.~\cite{Barbier:2004ez} reviews such interactions in the more general SUSY context.

\section{Dirac Gauginos}
\label{sec:dirac}
We have argued in the context of our 10 TeV effective SUSY theories that
naturalness requires sub-TeV gluinos, which provides a very
significant and visible SUSY production
channel at the LHC.  Yet, if we remain uncommitted to the structure of
physics above 1 TeV, we have argued that the gluino need not be
present in the sub-TeV effective theory. At first sight, these two
statements might seems in conflict, but in fact they merely exemplify
a general theme in SUSY models:  a very minimal field content in the
far IR often {\it requires} a less minimal field content at higher
energies. This is the case with regard to gauginos, and gluinos in
particular due to their stronger couplings. 
The idea of Dirac gauginos~\cite{Hall:1990hq,Randall:1992cq,Fox:2002bu} is to have extra field content in the form
of a chiral superfield, $\Phi_i$,  in the adjoint representation of each SM gauge
group, with soft SUSY breaking such that the $\Phi_i$ fermion, $\chi_i$, and the
gaugino, $\lambda_i$, get a Dirac mass with each other, $m_{\lambda_i} \lambda_i
\chi_i$. With such non-minimal field content below $10$ TeV we will
see that it is natural to have the Dirac gauginos heavier than $1$
TeV.

The 10 TeV effective theory with Dirac gauginos, analogous to
the construction of Eq.~\eqref{eqn:10tevlight}, is given by
\begin{eqnarray}
\label{eqn:10tevlightdirac}
{\cal L}_{eff} &=& \int d^4 \theta K + \left(\int  d^2 \theta \left(\frac{1}{4}{\cal W}_{\alpha}^2 + y_t \bar{T} H_u Q +  y_b \bar{B} H_d Q + \mu H_u H_d +(\sqrt{2}m_i\theta^{\alpha}){\cal W}_{i\alpha}\Phi_i \right)+{\rm h.c.}\right) 
\nonumber \\
&~& 
+ {\cal L}_{kin}^{light}
-  \left(\bar{u} Y_{u}^{light} h_u q_L
+ \bar{d} Y_{d}^{light} h_d q_L+{\rm h.c.} \right)+ {\cal L}_{lepton} \nonumber \\
&~&  - m_{\tilde{q}_L}^2 |\tilde{q}_L|^2  - m_{\tilde{t}^c_R}^2 |\tilde{t}^c_R|^2 - m_{\tilde{b}^c_R}^2 |\tilde{b}^c_R|^2  -m_{h_u}^2 |h_u|^2  - m_{h_d}^2 |h_d|^2 -m_{\phi_i}^2|\phi_i|^2\nonumber \\ &~&- B\mu h_u h_d - A_t \tilde{t}^c_R h_u \tilde{q}_L - A_b \tilde{b}^c_R h_d \tilde{q}_L+{\rm h.c.}\nonumber \\
&~& +{\cal L}_{hard} + {\cal L}_{non{\rm -}ren.},
\end{eqnarray}
where the explicit Grassmann $\theta^{\alpha}$ dependence parametrizes
the soft SUSY breaking Dirac gaugino mass term in superspace notation, and
$m_{\phi}^2$ in the third line gives soft mass-squared to the
scalars in the adjoint superfield $\Phi$. The remaining terms are as
discussed below Eq.~\eqref{eqn:10tevlight}. 

Similarly, the 10 TeV effective theory with Dirac gauginos, analogous to the
construction of  Eq.~(\ref{eqn:10tevheavy}), is given by 
\begin{eqnarray}\label{eqn:10tevheavydirac}
{\cal L}_{eff} &=& \int d^4 \theta K+
\left(\int  d^2 \theta \left(\frac{1}{4}{\cal W}_{\alpha}^2 + y_t \bar{T} H_u Q+(\sqrt{2}m_i\theta^{\alpha}){\cal W}_{i\alpha}\Phi_i \right)+{\rm h.c.}\right)\nonumber \\
&~& + {\cal L}_{kin} - \left(\bar{u} Y_{u}^{light} h_u q_L+y_b \bar{b} h_u^* q_L 
+ \bar{d} Y_{d}^{light} h_u^* q_L +{\rm h.c.}\right)+ {\cal L}_{lepton}
\nonumber \\
&~&  - m_{\tilde{q}_L}^2 |\tilde{q}_L|^2  - m_{\tilde{t}^c_R}^2 |\tilde{t}^c_R|^2  -m_{h_u}^2 |h_u|^2-m_{\phi_i}^2|\phi_i|^2
\nonumber \\
&~& - \left( A \tilde{t}^c_R h_u \tilde{q}_L + m_{\tilde{h}} \tilde{h}_u \tilde{h}_d + {\rm h.c.}\right)
\nonumber \\
&~& + {\cal L}_{hard} + {\cal L}_{non{\rm -}ren.}.
\end{eqnarray}
This scenario was first emphasized and studied in detail in the context of full supersymmetry in~\cite{Davies:2011mp}.

\subsection{Naturalness}

 Expanding the soft gaugino mass term from
superspace into components yields couplings,
\begin{equation}{\cal L} \supset \sqrt{2}m_{\lambda_i}D^i(\phi_i+
\bar{\phi}_i)-m_{\lambda_i}(\chi^i\lambda_i+\bar{\lambda}^i\bar{\chi}_i)
\end{equation}
The $D$-term contributes mass to the {\it real} part of $\phi_i$ so
that the total mass-squared is
$m_{R_i}^2 = 2(m_{\lambda_i}^2+m_{\phi_i}^2)$, while the imaginary
part has mass-squared of just  $m_{\phi}^2$. 
In addition, the $D$-term generates a coupling of the real part of
$\phi$ to the other scalars charged under the related gauge group. For
the case of Dirac gluinos, we obtain the coupling ${\cal L} \supset
-\sqrt{2}m_{\lambda_3}g_s(\phi_3^i+\bar{\phi}^a)(\bar{\tilde{q}}T^a\tilde{q})$,
where $T^a$ are the Gell-Mann color matrices. 
This provides a new correction to the stop mass-squared at one loop which
cancels the logarithmic divergence found in Eq.~\eqref{eqn:stopmass}~\cite{Fox:2002bu}. Eq.~\eqref{eqn:stopmass} is then
replaced by a UV-finite total correction, 
\begin{equation}\delta m_{\tilde{t}}^2= \frac{2g_s^2
    m_{\tilde{g}}^2}{3\pi^2}\ln\frac{m_{R_3}}{m_{\tilde{g}}}.
\end{equation}
Taking the stop much lighter than the gluino and the  
scalar gluon (``sgluon'') to be comparable to the gluino mass (the
above logarithm $\sim 1$), and requiring naturalness of the stop mass,
yields
\begin{equation}
m_{\tilde{g}} \lesssim 4 m_{\tilde{t}}.
\end{equation} 
This implies it is natural to have gluinos above a TeV for stops as
light as $\sim 300$ GeV. In such cases, it is sensible to remove
the gluino and sgluons from the sub-TeV effective theory, and from
early LHC phenomenology.

\subsection{R-parity violation}  
\label{sec:diracrpv}
As advertized in Subsection \ref{sec:rpvleptonnumber},  Dirac gauginos are also
 important for the case of lepton-number
conserving RPV because
they completely relax the stringent constraints from
neutron-antineutron oscillations by allowing one to have a $U(1)$
baryon number symmetry. The trick is that this symmetry is realized as
an R-symmetry in the sense that different fields in a supermultiplet carry different charges. The charges of the fields are given in table~\ref{tab:rch}. 
One can then check that Eq.~\eqref{eqn:10tevheavydirac} and the
RPV couplings of Eq.~\eqref{eqn:rpvudd} respect such a baryon number
$R$-symmetry in the absence of the $A$ term. 

\begin{table}[t]
\begin{center}
\begin{tabular}{|lr|lr|}
\hline
Boson & $q$ & Fermion & $q$ \\ \hline
\hline
$h_u$ & $0$ & $\tilde{h}_u$ & $-1$ \\ \hline
& & $\tilde{h}_d$ & $1$ \\ \hline
$\tilde{q}_L$ & $\frac{4}{3}$ & $q_L$ & $\frac{1}{3}$ \\ \hline
& & $(u_L,d_L)$, $(c_L,s_L)$ & $\frac{1}{3}$ \\ \hline
$\tilde{t}_R^c$ & $\frac{2}{3}$ & $t^c_R$ & $-\frac{1}{3}$ \\ \hline
& & $u_R^c$, $d_R^c$, $s_R^c$, $c_R^c$, $b_R^c$ & $-\frac{1}{3}$ \\ \hline
& & leptons & $0$ \\ \hline
$A_{\mu}$ & $0$ & $\lambda$ & $1$ \\ \hline
$\phi$ & 0 & $\chi$ & $-1$ \\ \hline
\end{tabular}
\end{center}
\caption{R-charges of particles in theory with Eq.~\eqref{eqn:rpvudd} and Dirac gaugino masses.}
\label{tab:rch}
\end{table}

With baryon R-symmetry, neutron-antineutron oscillations are forbidden,
even when RPV couplings are sizeable,  which raises the possibility
that stops can be singly produced at colliders.\footnote{Ref.~\cite{Kilic:2011sr}
  discusses a model in which it is $\tilde{b}_R$ that is singly
  produced (at the Tevatron), and in which neutron-antineutron
  oscillation placed important constraints. Dirac gauginos would also
  loosen these constraints in this context. (Our flavor estimates
  suggest that  $\tilde{b}_R$ lighter than TeV is disfavored, but
  perhaps this is possible with a more special flavor structure.)}
 But we first have to
ask if this is plausible in light of flavor physics and CP constraints. 
 A useful way to think of the new flavor structure of 
RPV couplings of $\tilde{t}^c_R$ in effective SUSY
 is that they effectively make this antisquark a ``diquark'', even up
 to its baryon number.  
In this way, the general discussion and constraints of flavor structure
for scalars with $d_R d_R$ diquark couplings given in~\cite{Giudice:2011ak} applies to the effective SUSY
setting here. In particular, Ref.~\cite{Giudice:2011ak} discusses the different plausible
hierarchical structures for such couplings and the mechanisms underlying their
safety from FCNC and CP-violating constraints. As is shown there, it
is indeed plausible for the  $\tilde{t}^c_R$ to have order one
couplings to light quarks, and therefore be singly produced.\footnote{A similar analysis is possible for (non-R-symmetry) baryon-number
preserving RPV and loosening the constraints from lepton-number
violation tests such as neutrinoless double-$\beta$ decay.}

Baryon-number R-symmetry, by forbidding the A-term, also makes for an
interesting signature for pair-production of $\tilde{q}_L$ since they
can no longer mix with $\tilde{t}_R^c$ after electroweak symmetry breaking. These
squarks do not directly couple to quark pairs, unlike $\tilde{t}_R^c$,
which means that each  $\tilde{q}_L$ will decay into two third generation quarks {\it
  plus} a quark pair. 

\subsection{Electric dipole moments}

With the baryon R-symmetry as described above, it is straightforward to check that 
all the soft SUSY breaking parameters can be made real by appropriate
rephasing of fields in  Eq.~\eqref{eqn:10tevheavydirac}. 
Therefore there are no new CP-violating contributions to electric
dipole moments from this Lagrangian. However, as discussed in Section
V, we should more realistically add third-generation flavor-changing corrections to
any such Lagrangian, which can contain new CP-violating
phases. However, as discussed there these new terms will be suppressed by 
${\cal O}(\epsilon^2)$. In this way, we expect non-vanishing but
highly suppressed new contributions to EDMs. These observations for
effective SUSY are closely related to the observations made in Refs.~\cite{Fox:2002bu,Hisano:2006mv,Kribs:2007ac}.

\section{Collider Phenomenology}\label{sec:pheno}
In this section we will demonstrate three things:
\begin{enumerate}
 \item After $\sim 1/$fb LHC running, there are analyses that put
   non-trivial constraints on the motivated parameter space of effective SUSY.
\item Nevertheless, very large parts of the parameter space, 
fully consistent with electroweak naturalness, are still alive.
\item  The most constraining searches for effective SUSY, so far, are not always those optimized for more standard SUSY scenarios.
\end{enumerate}

While effective SUSY has many interesting experimental regimes, we will not attempt a complete study in this paper. Rather, we will  focus on the simplest natural setting, and do enough of the related phenomenology to make the points (1 -- 3) above. 
We will pursue the R-parity conserving scenario for a few related reasons. This naturally provides the phenomenological handle of sizeable missing energy, which can stand out in even the early LHC data. Secondly, there is greater familiarity in  the community of pursuing these  event topologies. Our results for effective SUSY can then be compared with the phenomenology of more standard SUSY scenarios. In the paper, we have however emphasized that R-parity violation is a particularly well-motivated option within effective SUSY, and it does display distinctive phenomenological features. We will pursue a more detailed study of this kind of the RPV phenomenology of effective SUSY in future work.

The central consideration for effective SUSY phenomenology is the great reduction in new colored particles, squarks, compared with standard SUSY scenarios. 
In effective SUSY we keep just the minimal set of superpartners below TeV needed to stabilize the electroweak hierarchy.
This has the effect of lowering the new physics cross-sections substantially. 
Furthermore, in standard SUSY settings one typically entertains higher
superpartner masses than is technically natural, partly a result of
renormalization group running of super-spectra from very high scales,
and partly in order to radiatively raise the physical Higgs boson mass
above the experimental bound. In our bottom-up effective SUSY, with
less UV prejudice, we have only tried to constrain the spectrum from
the viewpoint of naturalness and the little hierarchy problem. As we
have seen, other mechanisms for raising the physical Higgs mass work
well within effective SUSY. Therefore, we favor  the regime where
stops are lighter than $500$ GeV, while gluinos may be so heavy as to be irrelevant  in the early LHC. The decay products of lighter stops in effective SUSY can easily fail to pass the harsher cuts on missing energy and jet energies used in searches optimized for heavier superpartners. 

In the following subsection, we will study in detail collider
constraints which one can put on the most minimal scenario, namely
light stops and sbottom (predominantly left-handed) with a neutralino
at the bottom of the spectrum. We will briefly review the Tevatron
constraints on this scenario and further analyze the constraints
arising from LHC data at $\cL \sim 1$ fb$^{-1}$. In the subsequent
subsection we will survey other variations, but will not go into  details. We leave this to future work.  

\subsection{Neutralino and Squarks}
\label{sec:minmodel}
In this particular subsystem we
 will simplify considerations even further to the effective theory 
of Eq.~\eqref{eqn:1tevlightneutralino}, where we have just a bino LSP lighter than the
squarks. The neutralino might more generally be an admixture of
several neutral gauge eigenstates, but phenomenologically this is not
very relevant; the neutralino  is simply
a way of invisibly carrying off odd R-parity from colored superpartner
decays. The bino is a good proxy
for such a general neutralino.  In the remainder of this section, we
focus on the collider phenomenology of Eq.~\eqref{eqn:1tevlightneutralino}. 

One further simplification we make is to take the stops and sbottom to
be roughly degenerate. If there is no substantial left-right mixing, this is a very good 
approximation in the left-handed (LH) sector. The mass difference between the LH stop and sbottom
is given by
\beq\label{tbdeltam}
\Delta m \approx \frac{m_W^2 \sin^2 \beta}{2 m_{{\tilde q}_L}}.
\eeq
Since this splitting comes from  $SU(2)\times U(1)$ D-terms, it is
proportional to the mass of the $W$.
Usually if the splitting is dominated by D-terms, one gets that
$m_{\tilde b} > m_{\tilde t}$. This might suggest that one should also consider a decay mode 
$\tilde b \to W^{(*)} \tilde t$. However this would imply a three-body
decay, which is  therefore highly suppressed. More important, stop decay modes $\tilde t \to W^{(*)}  \tilde b $ can become 
competetive  to other stop decay modes, if it is forced to proceed through an off-shell top.
However this can happen only if the left-right mixing between the stops is large, and we will neglect 
this possibility further.    

Before considering the LHC,  we should note several D0 searches which directly address this 
scenario. The first relevant search looks for b-jets $+ \slashed E_T$ ~\cite{Abazov:2010wq}. 
This search constrains the sbottom mass to be higher than 247 GeV if the neutralino is 
massless. The constraints become weaker if the neutralino is heaver, but unless there is 
an accidental degeneracy, the lower bounds on the sbottom are still around 200 GeV. Another search of D0 looks 
for stops, which are pair-produced and further decay into $b\ l +
\slashed E_T$ (where this decay mode is 
assumed to have 100 \% branching fraction). The most updated search used events with opposite \
flavor pairs~\cite{Abazov:2010xm}. This search also bounds the stop mass at 240 GeV if the 
neutralino is massless and for massive neutralino (without any accidental degeneracy with 
the stop) the bound is of order 200 GeV, depending on the neutralino mass.

CDF has a more elaborate search, where it looks for $t\bar t + \slashed E_T$. This search was 
performed in monoleptonic~\cite{Aaltonen:2011rr} and
hadronic~\cite{Aaltonen:2011na} channels. The bounds one can put on
production cross sections from theses two  measurements are comparable
to each other, but too weak to constrain effective SUSY with its small 
 squark cross section.\footnote{Hereafter we do not consider a mass range of stop below 200 GeV, where the stop mostly decays off-shell. This intriguing possibility is not yet excluded, and the reader is refered to~\cite{Kats:2011it,Kats:2011qh}.}   

Now let us turn our attention to the LHC searches. As we will see, the
bounds from the LHC are not very stringent (partly due to an
insufficient number of dedicated searches). This is in part because, 
with the exception of  an Atlas top-group search for $t\bar t +
\slashed E_T$ (which we will discuss later), there are no dedicated
searches for this scenario. However there are several general
searches, which can be sensitive to the stop/sbottom/neutralino
subsystem we are studying here.
We explictly considered the following list of searches:
\begin{enumerate}
 \item jets + $\slashed E_T$ (including simple $\slashed H_T$ search and an $\alpha_T$ search)~\cite{Chatrchyan:2011zy,CMS:missht} 
 \item jets + $\slashed E_T$ with b-tag~\cite{CMS:btag,Atlas:btag}
 \item lepton + jets + $\slashed E_T$~\cite{CMS:monolep}
 \item OS dileptons + jets + $\slashed E_T$~\cite{CMS:osdilep}
 \item lepton + jets with b-tag + $\slashed E_T$~\cite{Atlas:blmet}
\end{enumerate}

In order to estimate the bounds on our scenario, we simulated events
and checked the acceptances within the channels listed
above.\footnote{Whenever both Atlas and CMS have performed closely
  overlapping searches, we have considered just the CMS representative.
The relevant Atlas searches are~\cite{Collaboration:2011iu,Aad:2011ib}.
We also did not explicitly simulate an additional CMS jets $+ \slashed
E_T$ search which takes advantage of the $m_{t2}$
variable~\cite{CMS:mt2}, since it is not expected to have a good
acceptance in our case. } The events were generated and decayed with
{\tt MadGraph 5}~\cite{Alwall:2011uj} and further showered and
hadronized with {\tt Pythia 6}~\cite{pythiamanual}. The events were
reconstructed with {\tt FastJet-2.4.4}~\cite{Cacciari:2005hq}. We
calculated 
all the NLO cross-sections with {\tt Prospino 2}~\cite{Beenakker:1996ed} and reweighted
 all the events appropriately. We ran
each spectrum assuming that the mass difference between the stops and
sbottom are negligible. Given the mass difference, Eq. \eqref{tbdeltam}, this is not a bad approximation. (One can of course play with the mass difference between $\tilde t_L $ and $\tilde t_R$, still keeping the spectrum natural, but we did not perform this study.)
\begin{figure}[t]
\centering
\includegraphics[width=11.7cm]{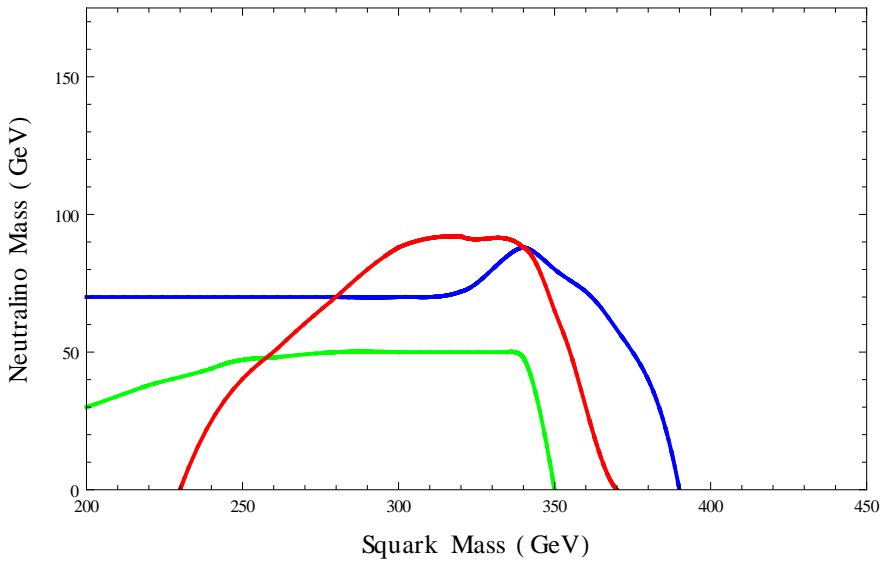}
\caption{Exclusion curves for our minimal model, Eq.~\eqref{eqn:1tevlightneutralino},
from three relevant searches as a function
  of masses for squarks and neutralino. We assume roughly equal masses
  for all three squark
  species, two stops and a sbottom. The green line represents exclusion by $\alpha_T$ search, the blue line is an exclusion by $\slashed H_T$ search and the red one is exclusion by $t\bar t + \slashed E_T$ search.}
\label{fig:allbounds}
\end{figure}   

\begin{figure}[t]
 \centering 
\includegraphics[width=11.7cm]{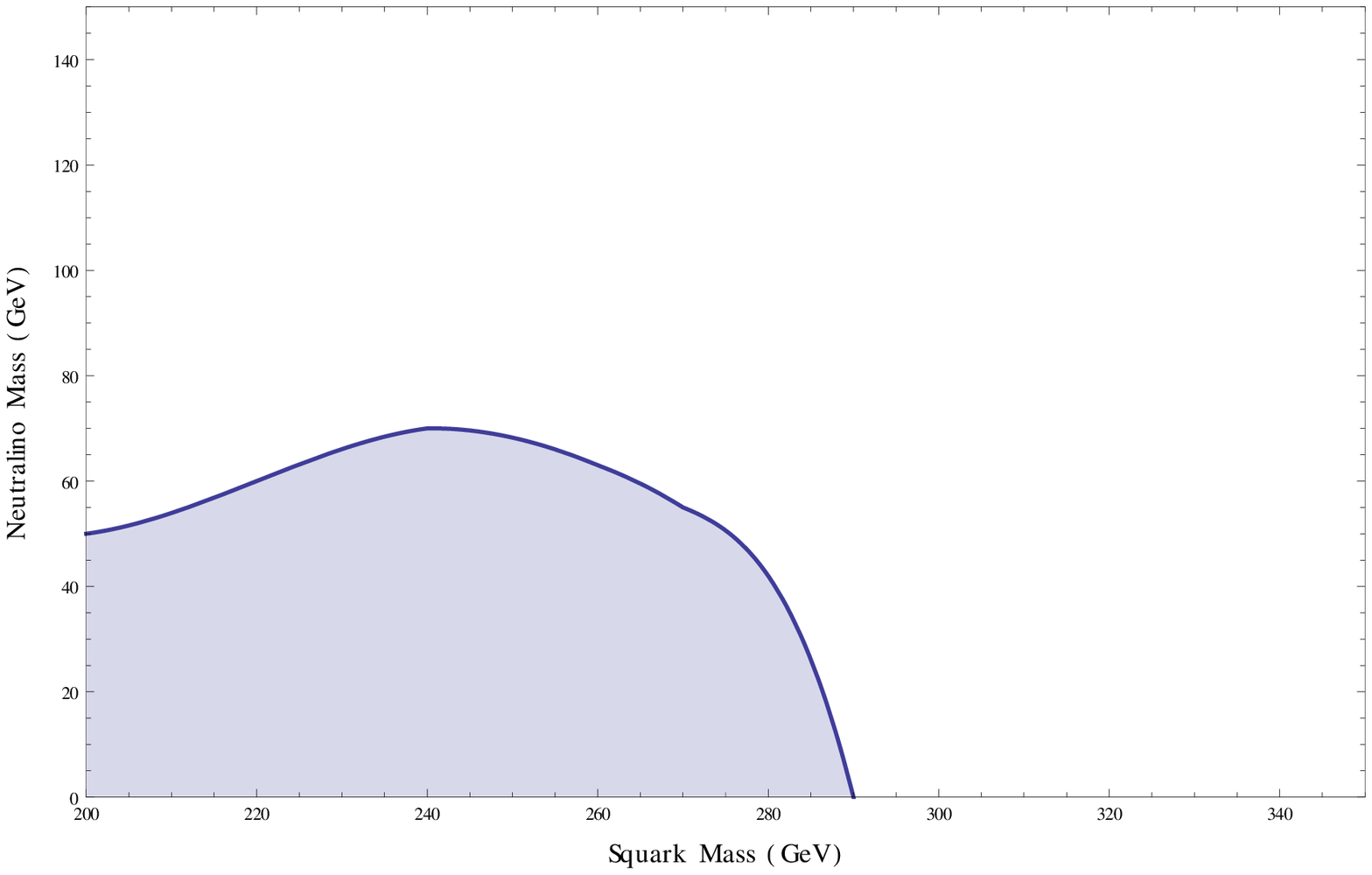}
\caption{Exclusion of a single sbottom due to jets + $\slashed H_T$ search as a function of a sbottom and neutralino masses.}
\label{fig:singsb}
\end{figure}

 We find that all the searches listed above, except searches for jets +
$\slashed E_T$, do not put any interesting bounds on the subsystem
that we are  discussing here. The searches in leptonic modes put
extremely harsh cuts on the $H_T$ of the entire event, and therefore
easily miss the stops in the range between 200 and 400 GeV, while the
cross sections in the higher mass range are far too
small. Unfortunately, the Atlas search for  jets $+ l +  b\rm{-tag} +
\slashed E_T$ ~\cite{Atlas:blmet} also does not add interesting
constraints, mostly because it is tuned to detect (or exclude) gluinos
above 400 GeV which further cascade-decay to bottom, top and
neutralino.\footnote{This search claims that it looks for events with
  4 b-jets with lepton and $\slashed E_T$, however demands only a
  single b-tag in the event selection. One can probably put more
  interesting bounds by demanding more than one b-tag.} The jets +
$\slashed E_T$ searches indeed put interesting constraints on our
stop/sbottom/neutralino subsystem and we show our bounds 
in Fig.~\ref{fig:allbounds}.
 We found that more than half of all the
relevant events which contribute to the exclusion come from sbottom
production and decays. In fact, even a single 
sbottom without any stops would be excluded all the way to 300 GeV with the
same searches for massless neutralino. For more
general neutralino mass the
single-sbottom 
 exclusion plot appears in Fig.~\ref{fig:singsb}. 
By comparison, the same searches put no bounds on a single stop (or
even both stop species), due to extremely bad acceptance in this range of masses.  

\begin{figure}[t]
\centering
\includegraphics[width=11.7cm]{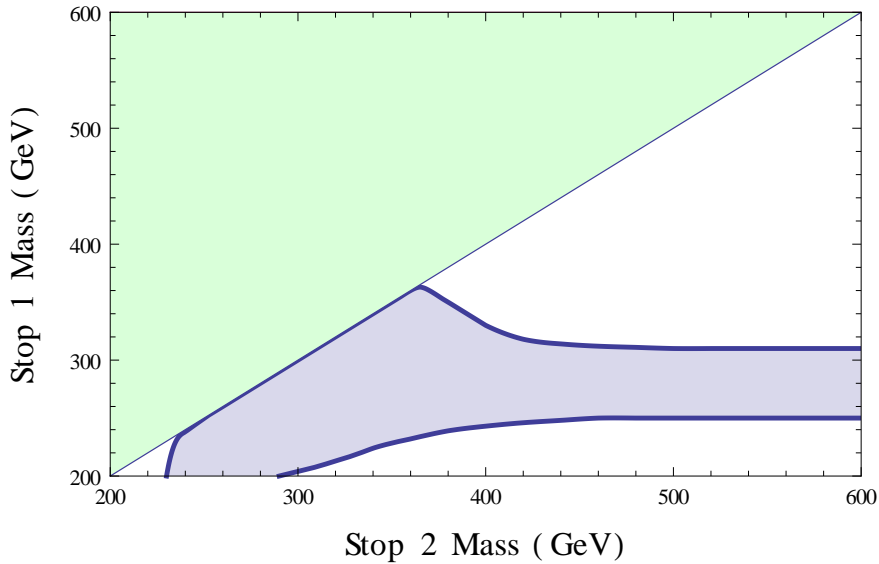}
\caption{Exclusion curves for two stops with different masses from the Atlas search for $t\bar t +\slashed E_T $ in monoleptonic channel~\cite{Aad:2011wc} . The neutralino mass is assumed to be zero. Note a narrow band between 250 and 290 GeV for the first stop which is excluded even when the second stop is very heavy. This is the region where the sensitivity of the search is maximized.}
\label{fig:stops}
\end{figure}

This, however, does not conclude the full list of searches. There is
an additional search by Atlas, which looks precisely for $t \bar t
+\slashed E_T$ in a monoleptonic channel~\cite{Aad:2011wc}. This
particular search puts almost no bound for production of a single
species of stop, but the picture is different when we have both stops
roughly degenerate (with double the production cross sections). We
show the final exclusion plots on Fig. \ref{fig:allbounds}, where the
exclusion due to $t \bar t + \slashed E_T$ search is given by the red
curve. On Fig.~\ref{fig:stops} we show the ranges excluded by this search if we split the masses of the stops (neutralino mass is assumed to be zero). Note that this exclusion is comparable to the exclusion one
gets with the jets $+\slashed H_T$ search.  

\subsection{Overview of some other possibilities}
\subsubsection{Gluinos}
\label{sec:phenogluinos}

Because of their large color charge and the high multiplicity of their
decay products, the biggest phenomenological consideration for the 7
TeV LHC is the presence or absence of gluinos below a TeV. 
Production cross-section grows significantly as gluinos are taken below 1 TeV in
mass, and gluinos decay exclusively into the third generation
squarks. This scenario has been studied both in cases when the gluino
decays into a sbottom (see abovementioned searches for jets plus
$\slashed E_T$ with a b-tag) or into a
stop~\cite{Atlas:blmet}. However, there are reasons to believe that a
monoleptonic channel with one b-tag, which was used in the Atlas
search is not  optimal.  The model of gluinos decaying exclusively to
stops was carefully studied in~\cite{Kane:2011zd} and it was found
that with luminosity of 1 fb$^{-1}$ gluinos up to 650 GeV can be
discovered, if one takes advantage of a few competitive channels, like same-sign dileptons, multileptons with or without b-tags (and sometimes multiple b-tags).  

\subsubsection{Collider-stable squarks}
\label{sec:phenostablesquark}
One can also consider the very simple scenario  with $\tilde t_L,\ \tilde t_R $ and $\tilde b_L$ 
 at the bottom of the superpartner spectrum. With
 R-parity, the lightest scalar (either stop or sbottom) is stable. We
 should of course assume that it decays at some point (for example it
 can decay into a gravitino, or through some tiny R-parity violating
 coupling) in order to avoid constraints from searches for ultra-heavy
 hydrogen atoms~\cite{Yamagata:1993jq}, but this still allows  squarks with 
 cosmological lifetimes~\cite{Kang:2006yd}. If this is the case, $\tilde
 t$ or $\tilde b$ should show up as R-hadrons at the LHC. Recent
 bounds from CMS impose severe constraints on this scenario if the
 lightest superpartner  is a stop~\cite{CMS:Rhads}.\footnote{Even though
   the authors of this paper do not interpret there results in terms
   of stable $\tilde b$, there is no reason to believe that this bound
   would be dramatically different.} Results of these searches imply
 that a stable stop in the mass range between 100 and 800 GeV is excluded
 if its production cross section is of order $10^{-2}$ pb. Comparing
 these results to theoretically expected production cross
 sections~\cite{Beenakker:2010nq}, we find that these cross-sections
 are expected for a single stop with mass up to 600 GeV. However in
 our case, we should at least multiply the cross sections by a factor of
 three (we have two stops and at least one sbottom), rendering the
 bound to somewhat higher than 600 GeV. Therefore, if one takes the
 little hierarchy problem seriously up to $\sim 10$ TeV,  this
 scenario is disfavored.\footnote{However, as noted in
   Subsection~\ref{sec:1tevheavy}, 
 the effective theory of Eq.~(\ref{eqn:1tevheavy}) is a
useful departure point for adding in RPV phenomenology.}

\subsubsection{Neutralino and Chargino LSPs}

A safer option is to consider the effective theory of Eq.~(\ref{eqn:1tevlightneutralino}), where we
see the Higgsinos providing natural neutralino/chargino candidates. If
the neutralino is the LSP, bounds on stable charged or colored
particles are evaded. Of course, the neutralinos and charginos may
more generally be an admixture of several electroweak gauge
eigenstates.

In detail, the presence of a chargino as an NLSP makes a
phenomenological difference, but we believe that it is less decisive
in the present context. The difference from the scenario described in
Subsection~\ref{sec:minmodel} is that on top of the decay modes $\tilde t
\to t \tilde \chi^0$ and $\tilde b \to b \tilde \chi^0$ we have
already considered, we will have competing modes $\tilde b \to t
\tilde{\chi}^\pm$ and $\tilde t \to b \tilde{\chi}^\pm$. Since we are mostly
interested in the region of mass parameters where the top-quark mass
is far from negligible, we conclude that the decay mode $\tilde b \to
t \tilde{\chi}^\pm$ will be mostly suppressed due to the phase
space. Therefore, introducing the chargino at the bottom of the
spectrum will usually have a mild effect on sbottom decay modes and
the constraints which come from these decays (mostly jets plus
$\slashed E_T$). However the stops decay modes will be altered
compared to our discussion in Subsection~\ref{sec:minmodel}, since the
decay mode $\tilde t \to b \tilde \chi^\pm$ is now phase space
unsuppressed.  The chargino will consequently decay to the neutralino
and $W$  (maybe off-shell). Therefore,  this will look roughly
similar to the decay modes of a regular stop, even though the
kinematics might be different. If the chargino and neutralino are
quasi-degenerate, then the decay modes of stops very much resemble
those of sbottoms, thereby effectively increasing the production cross
sections for jets plus $\slashed E_T$ and making the constraints
somewhat more stringent then what we find in Subsection~\ref{sec:minmodel}.  

While the above are reasonable deductions, explicit simulation is
still required when charginos are light. We again leave this to future work.
 
\section{Outlook}\label{sec:future}

In this paper, we developed a bottom-up formulation of
 effective supersymmetry  and analysed some of its phenomenological
 aspects. As we have shown, the constraints on effective SUSY, even
 with the most conservative approach, 
R-parity with neutralino at the bottom of the spectrum, are very
mild. 
With these assumptions, the data still allow a spectrum fully
consistent  with electroweak naturalness.

This conclusion strongly suggests the future research program in this
direction. Evidently, current LHC searches are not optimized for this
scenario. It would be interesting to see how one can increase the
sensitivity of the current searches and vary the cuts so as to
allow better acceptance for effective SUSY. 
We expect that there is a strong opportunity for
 searches optimized to effective SUSY to make great inroads into
 discovery or exclusions within $\sim 10/$fb of LHC running, in the coming year.

Another promising avenue one can take has to do with R-parity violation. As we
emphasized in Section~\ref{sec:rpv}, RPV is highly motivated if
effective SUSY indeed describes the physics immediately beyond the
SM. Even the signals of RPV SUSY with lepton-number violation can be
quite
 challenging if squark decays into leptons involve $\tau$.
The signals of RPV SUSY with baryon-number violation are even more
challenging,
 because the decays of the squarks will mostly results in
 jets. However, as pointed out for the case with baryon R-symmetry, 
squarks can have more spectacular decays into several jets, including
two with heavy flavor.
Current exotica searches~\cite{Chatrchyan:2011cj} put very mild
 bounds on these RPV scenarios and it is very interesting if one can improve these search strategies to get better sensitivity to the new physics.       

\acknowledgments{We are grateful to Nima Arkani-Hamed, Nick Hadley,
  Greg Landsberg, Patrick Meade, Rabi Mohapatra, Michele Papucci, Matt Reece, Paolo
  Rumeira, Yael Shadmi for useful discussions. AK is partially supported by NSF grants
  PHY-0855591 and PHY-0801323 . CB and RS are supported by NSF under grant PHY-0910467
  and by the Maryland Center for Fundamental Physics. AK and RS are
  very grateful to the Aspen Center for Physics where some of this research was done.}


\bibliography{lit}
\bibliographystyle{apsper}

\end{document}